\documentclass[%
reprint,
 amsmath,amssymb,
pra,
]{revtex4-1}

\usepackage{graphicx}
\usepackage{dcolumn}
\usepackage{bm}
\usepackage{subfigure}
\usepackage{physics}
\usepackage[space]{grffile}
\usepackage{color}
\usepackage{empheq}

\begin{document}


\title{Light-cone and local front dynamics of a single particle extended quantum walk}

\author{Hemlata Bhandari}

\author{P. Durganandini}
 
\affiliation{%
Department of Physics, Savitribai Phule Pune University, Pune 411007, India\\
}%

\date{\today}
            
\begin{abstract}

We study the light-cone and front dynamics of a single particle continuous time extended quantum walk on a one dimensional lattice with finite range hopping. We show that, in general, for an initially localized state,  propagating wave fronts can be characterized as ordinary or extremal fronts with the latter exhibiting an anomalous sub-diffusive scaling behaviour in the front region.  We investigate the dynamical global and local scaling properties of the cumulative probability distribution function for the extended walk with nearest and next-nearest neighbour hopping using analytical and numerical methods. The global scaling shows the existence of a 'causal light-cone'  corresponding to excitations travelling with a velocity smaller than a maximal  'light velocity'.  Maximal fronts moving with fixed 'light velocity' bound the causal cone. The front regions  spread with time sub-diffusively exhibiting  a local Airy scaling which leads to an internal staircase structure.  At a certain critical next-nearest neighbour hopping strength, there is a transition from a phase with one 'causal cone' to a phase with two  nested 'causal cones' and the existence of an internal staircase structure in the corresponding cumulative distribution profiles.  We also connect the study to that in spin chain systems and indicate that a single particle quantum walk on the  one dimensional lattice already captures the many body physics of a spin chain system. In particular, we suggest that the time evolution of a single particle quantum walk on the one dimensional lattice with an initially localized state is  equivalent to the time evolution of a domain wall initial state in a corresponding spin chain system.

\end{abstract}

\maketitle

\section{\label{sec:Intro}Introduction}

 The study of quantum walks has become a subject of great interest in recent years \citep{Aharonov,Kempe,VAndraca}. They have been used to model a variety of diverse physical processes like that of coherent transport of excitations between different potential sites \citep{mulken,longhi}, relativistic effects like Zitterberwegung and Klein tunnelling \citep{QW_Zitterberwegung, QW_Klein_tunneling}, Anderson localization \citep{QW_Anderson_localization}, topological phases \citep{QW_Topological_phases}, strongly correlated many body systems \citep{QW_strong,Nata_Andrei}, etc.  These studies have gained further impetus due to experimental realizations of quantum walks in systems of  trapped ions \citep{QW_trapped_ion, PRA.65.032310,PRL.103.090504}, cavity QED \citep{QW_QED_PhysRevA.67.042305}, photon waveguide arrays \citep{QW_photons}, ultra-cold atoms in optical lattices \citep{QW_optical_lattice, QW_trapped_atom, QW_BEC}, etc.  
 
 Quantum mechanical notions of coherence, superposition and interference lead to very different behaviour for quantum walks as compared to that for classical random walks.  Studies on single particle continuous time  quantum walks modelling tight binding Hamiltonians with nearest and next-nearest neighbour hopping on one dimensional lattices  show that the walk (for an initial localized state) typically displays ballistic front propagation instead of the diffusive nature of classical random walks \citep{Toro, QW_Krapivsky1,QW_Krapivsky2}. The spread of the wave packet is bounded by the existence of a maximal group velocity with which the fronts can propagate. The local probability exhibits an anomalous sub-diffusive scaling in the front region \citep{Toro,QW_Krapivsky2}. Existence of a maximal velocity at which information can spread or correlations can develop has long been known in the context of non-relativistic spin systems \citep{lieb-robinson}. The Lieb-Robinson(LR) theorem asserts that in any non-relativistic quantum spin system governed by a  Hamiltonian with short range interactions,  the velocity with which information or physical effects can  propagate in the system is bounded. The existence of such a velocity bound  has been shown to lead to a causal structure with connected correlations existing only within a 'light-cone' even in non-relativistic systems \citep{lieb-robinson,bravyi2006, cardy2006}. The emergence of light-cone effects has also been shown in the context of the time propagation of  initial domain wall magnetization states in an Ising spin chain in a magnetic field \citep{antal_pre59, sasvari_pre69}.  The magnetization profiles exhibit a global scaling and a local scaling different from that in the bulk near the propagating fronts. The local scaling near the fronts was also shown to lead to the existence of an internal staircase structure near the front \citep{sasvari_pre69}.  In recent years,  investigations of Lieb-Robinson bounds and related light-cone effects have become a key issue in the study of non-equilibrium dynamics after global quenches in spin systems \citep{bonnes2014}. The physical interpretation is based on the idea that in a quench problem, the initial state acts as a source of pairs of entangled quasiparticles wherein quasiparticles having opposite momenta move ballistically.  Light-cone effects after quenches have also been demonstrated experimentally in ultra cold atoms \citep{cheneau2012}.  Most of these studies have been in the context of spin chains and many body systems \citep{Krapivsky3,collura,najafi2018} which are typically modelled in terms of tight binding Hamiltonians.  Since continuous time quantum walk systems provide an alternate view of tight binding Hamiltonians, it would be interesting to study light-cone and front dynamics in a quantum walk system.

       In this work, using analytic and numerical methods, we study the time dynamics of a single particle extended quantum walk (with the particle allowed to hop up to a finite number of neighbouring sites) with a view to investigate the light-cone and front  dynamics.  We show that, in general, for an initially localized state, propagating wave fronts can be characterized by an integer $k( \geq 0)$. At large times, near a $k$-th order front, the probability density scales as $1/t^{2/k+2}$. Hence, near a $k$-th order front with $k>1$, one observes sub-diffusive behaviour.  We obtain the dynamical global and local scaling properties of the cumulative probability distribution function for a quantum walk (QW) with nearest (NN) and next-nearest neighbour (NNN) hopping. The local probabilities show ballistic front propagation with additional internal fronts emerging when the next-nearest neighbour hopping exceeds a certain critical value \cite{QW_Krapivsky2}. We show that the global scaling of the cumulative distribution function exhibits the existence of a 'causal light-cone' inside which excitations travel with a velocity smaller than a maximal 'light velocity'. Maximal extremal fronts moving with fixed 'light velocity' bound the causal cone. The front regions spread with time sub-diffusively exhibiting a local Airy scaling. The latter is shown to lead to an internal staircase structure which we interpret as due to the quantization of particle number.  The cumulative distribution functions exhibit two different global and local scaling behaviours and the emergence of an internal staircase structure when the NNN hopping strength exceeds the critical value. It indicates a transition from a phase with one 'causal cone' to a phase with two nested 'causal cones'.  The emergence of nested causal cones and staircase structures suggests the existence of two different 'velocities of light' and two kinds of propagating quasiparticles.  We have also studied the behaviour of various localization measures like the Shannon entropy, inverse participation ratio and the return probability which point to a localization transition at the critical NNN hopping strength separating the two phases. 
                     
                      Further, we  relate the study to that of light-cone and front  dynamics in spin chain systems.  Specifically, for the simple QW (with only NN hopping), we observe that the time evolved cumulative distribution profiles for an initial spatially localized particle correspond exactly to the magnetization profiles of time-evolved domain wall states in the Ising chain \citep{antal_pre59, sasvari_pre69}.  We note that the Ising chain Hamiltonian can be transformed to a NN tight binding (TB) Hamiltonian by a Jordan-Wigner transformation \citep{Jordan-wigner}.  Also, the simple QW with only NN hopping is an alternative view of the NN TB Hamiltonian. This indicates that the time evolution of a single particle QW on the one dimensional lattice with an initial spatially localized state  is equivalent to the time evolution of a domain wall initial state in the corresponding spin chain system.  We extend this comparison further and suggest that the cumulative distribution profiles for the QW model with NN and NNN hopping with an initially localized state correspond to the magnetization profiles of an initial domain wall state time-evolving according to the integrable spin chain Hamiltonian  with three spin interactions of the XZX + YZY type \citep{suzuki,*suzuki2}. The observed two phases in the QW model with NN and NNN hopping can also be connected to the two different phases in the above integrable spin chain model \citep{titvinidze}.  The study provides an example of an integrable  model  showing the emergence of internal ‘light-cones’  with the existence of more than one kind of propagating quasiparticle. 
                      
  The plan of the paper is as follows. In Section~\ref{sec:model}, we describe the long time dynamics of a single particle extended quantum walk  using the saddle point approximation. We show that, at large times, for an initially localized state, propagating wave fronts can be characterized as ordinary or extremal fronts with the latter exhibiting an anomalous sub-diffusive scaling behaviour in the front region. In Section~\ref{sec:global}, we describe the light-cone dynamics and global scaling of the cumulative probability distribution for a QW with nearest and next-nearest neighbour hopping using analytic and numerical methods. We discuss in Section~\ref{sec:staircase}, the local scaling and staircase properties near the extremal front regions. We also discuss in Section~\ref{sec:global} and Section~\ref{sec:staircase}, the connections of the cumulative probability distributions with the time evolved magnetization profiles in spin chain systems. In Section ~\ref{sec:localization}, we study the localization dynamics of the wavefunction. We conclude with a brief summary and discussion of our results in Section ~\ref{sec:conclusions}.

\section{\label{sec:model} Extended quantum walk model: Long-time dynamics}
  The Hamiltonian for the continuous time quantum walk of a single particle on an one-dimensional lattice with hopping allowed between neighbours separated up to a finite range $M \geq 1$ ($M$, finite), can be written in the second quantized form as:  
\begin{equation}\label{eq:Hamiltonian}
H = \sum_{J=1}^M g_{J}\,\,\left(\sum_{n=-N}^{N‎}({c_{n+J}^{\dagger}(t)c_n(t)+ c_{n}^{\dagger}(t)c_{n+J}(t)‎})\right)
\end{equation}    
where $c_n^{\dagger}$, $c_n$ denote the usual creation and annihilation operators. We assume here a lattice with $2N+1$ sites with lattice spacing $a$. 
$g_J$ denotes the hopping amplitude between neighbours separated by a distance $J\quad(1 \leq J \leq M)$. In the following and the rest of the paper, we measure energy in units of the nearest neighbour coupling strength $g_1$ ($\hbar$ has been set to $1$.) Time is measured in units of $1/g_1$, length is measured in units of the lattice spacing $a$ and velocity is measured in units of $ag_1$. The probability density is measured in units of $1/a$. The single particle wavefunction $\psi(n,t)$ at the $n$-th site at time $t$ is obtained from the field operator $\Psi(t) = \sum_n c_n| n \rangle$ as $\psi(n,t) = \langle n |\Psi (t)\rangle$. Here $|n\rangle$ denotes the single particle position space eigen-basis vectors. From Eq.~\ref{eq:Hamiltonian}, it is easy to see that the single particle position space wavefunction satisfies the equation of motion (we set $\hbar =1$): 
\begin{equation}\label{eq:Eq_of_motion}
i\frac{d\psi(n,t)}{dt}= \sum_{J=1}^M g_{J}\,\,[\psi(n+J,t)+\psi(n-J,t)] \\
\end{equation}
We can solve the above equation of motion by a Fourier transformation to the momentum space. The momentum space eigenfunctions are plane waves: $\psi_n(q) = e^{i q n}$ where $q$ is the wave-vector measured in units of $1/a$. We set $a=1$ in the rest of the paper. The single particle energies $\omega(q)$ are given by the dispersion relation:
 \begin{equation}
w(q)= \sum_{J=1}^M 2g_{J} \cos (J q) 
\label{eq:dispersion}
\end{equation}

The wavefunction at the site $n$ and at time $t$ can be obtained as the Fourier sum over all wave-vectors $q$ lying in the first Brillouin zone: 
\begin{equation}\label{eq:wavefunction_sum}
\psi(n,t) = \frac{1}{L}\sum_q  e^{i( nq- w(q)t)}\hat \psi(q,0);\qquad -\pi < q \leq \pi
\end{equation} 
where $\hat \psi(q,0)$ denotes the initial wavefunction in momentum space at time $t=0$ and $L=2Na$. Consider now a particle initially localized at the site $0$, that is, $\psi(n,t=0)=\delta_{n,0}/\sqrt{L}$ (and therefore $\hat\psi(q,0)=\sqrt{L}$). In the limit of an infinite site lattice, the summation over the wave-vectors $q$ can be converted into an integral and the wavefunction at the $n$-th site  obtained in terms of the momentum integral over the first Brillouin zone:
\begin{equation}\label{eq:fourier_integral}
\psi(n,t) = \int_{-\pi}^{\pi} \frac{dq}{2 \pi} e^{i( nq- w(q)t)}\hat \psi(q,0)
\end{equation}
We note that  the Hamiltonian (Eq.~\ref{eq:Hamiltonian}) is symmetric under the reflection transformation: $n\rightarrow -n$; 
hence $\psi(-n,t) = \psi(n,t)$. 

When only nearest neighbour hopping is considered, the single particle energies are given as: $w(q)=2 \,g_1\,\cos q$. The Fourier integral in Eq.~\ref{eq:fourier_integral} can be then performed exactly and the wavefunction at the $n$-th site obtained in terms of the $n$-th order Bessel functions as~\cite{QW_Krapivsky1,Toro,cuevas}: $\psi(n,t)=i^{-n}J_n(v_{e}t)$.  The argument of the Bessel functions is proportional to the time $t$, with proportionality factor $v_e$, the maximum of the group velocity, $v_e = \max |v_g|= |\frac{ d\omega(q)}{ dq}|= 2 g_1$. In the presence of higher range hopping, that is, for $g_J\neq 0, J \geq 2$, the integral, Eq.~\ref{eq:fourier_integral}, cannot be solved exactly.  We need to resort to numerical evaluation of the integral. However, we can evaluate the integral for asymptotically large times ($t \rightarrow \infty$) by using the saddle point approximation. This allows us to get insight about the  behaviour of the wavefunction at large times. To this end, we write Eq.~\ref{eq:fourier_integral} as~\citep{Toro, QW_Krapivsky2}:
\begin{equation}\label{eq:saddle-point}
\psi(n,t) = \int_{-\pi} ^{\pi} \frac{dq}{2 \pi} e^{i\varphi(n ,q) t}; \quad  \varphi(n,q)\equiv\frac{n}{t}q-w(q)
\end{equation}
We note that under site  reflection symmetry, $\varphi(-n,q) = \varphi(n,-q)$. The saddle point approximation assumes that the dominant contribution to the integral comes from a small region of $q$ around the saddle point solutions $q^*$ satisfying the equation \citep{Toro, QW_Krapivsky2}:
\begin{equation} \label{eq:saddle_pt_equation}
\varphi'(n, q^*) = 0 \implies \omega '(q_n^*) =v(q_n^*)= \frac{n}{t} 
\end{equation}
where $f'(q)\equiv\frac{df(q)}{dq}$.
The integral in Eq.~\ref{eq:saddle-point} is then performed by expanding  $\varphi(n,q)$ around the saddle point solutions $q_n^*$ and summing over all the saddle point solutions $q_n^*$.

  It is evident from Eq.~\ref{eq:saddle_pt_equation} that the saddle point solutions describe ballistic propagation of left and right moving wave fronts travelling with group velocity $v(q)= \pm |\omega '(q)|$. For any bounded dispersion relation of the kind given in Eq.~\ref{eq:dispersion}, the group velocities $v(q)$ are bounded: 
  \begin{equation}
 -|v_{e}(q)| \leq v(q) \leq |v_{e}(q)|; \qquad v_e(q) = \mbox{max}|(w'(q))|. 
  \end{equation} 
  For $|n|<n_e(=v_e t)$, the solutions $q_n^*$ of the saddle point equation~(Eq.~\ref{eq:saddle_pt_equation}) are real,  leading to oscillatory solutions for the wavefunction which decay with time according to a power law behaviour.  When $|n|>n_e$, the solutions $q_n^*$ are imaginary, leading to exponentially decaying wavefunctions. Thus, the wave packet spreads with time with the propagation bounded by the maximal group velocity with which the fronts can travel. The maximal spread of the wave packet at any given instant of time is hence given by $2 n_e = 2 v_{e} t$.   
  
   In the following, we  classify fronts of order $k$ as that for which the first non-zero derivative of $v(q)$ ($q$ is real) occurs at order $k+1$($k \geq 0$), i.e, 
\begin{equation}\label{eq:front}
v^{(1)}(q) = 0 = v^{(2)}(q) = \cdots = v^{(k)}(q);\quad v^{(k+1)}(q) \neq 0
\end{equation}	
where $f^{(n)}(q) \equiv \frac{d^nf(q)}{dq^n}$.
Note that for  fronts of order $k \geq 1$, the first derivative of the front velocity or equivalently, the second derivative of the energy dispersion $\omega(q)$ is identically zero. We shall call the zero-th order front as an ordinary front and higher order fronts as extremal fronts. We note that the maximal fronts, i.e, fronts travelling with the maximal group velocity, $\pm v_e$ are first order extremal fronts. Additional extremal fronts can occur if there exist  
extremal solutions of Eq.~\ref{eq:front} at real $q$ values other than the maximal values $\pm q_e$.
 
    Near a $k$-th order front, and for asymptotically long times, the  integral in Eq.~\ref{eq:saddle-point} can be evaluated by expanding $\varphi(n,q)$  about the saddle point solution $q_n^*$ as :
\begin{equation}
\varphi(n,q)= \sum_{m=0}^{\infty}(q-q_n^*)^m \frac{\varphi^{(m)}(n, q_n^*)}{m!} 
\end{equation}
The wavefunction at site $n$ is then obtained as (keeping terms upto the $k+2$-th order in the expansion of $\varphi(n,q)$):
\begin{center}
\begin{eqnarray}\label{eq:front_solution}
\psi(n,t) & \approx &‎e^{i t \varphi(n,q_n^{*})}\int_{-\pi} ^{\pi}\frac{dq}{2 \pi} e^{-i t\frac{(q-q_n^{*})^{k+2}}{2!}\omega^{(k+2)}(q_n^{*})} \nonumber \\
&  \approx & e^{i \phi(n,q^*)t}\,\, e^{-i \frac{\pi}{2}{\frac{1}{k+2}}} \,\,\frac{\Gamma(\frac{1}{k+2})}{2 \pi (k+2) \,\rho } \,\, \frac{1}{t^{\frac{1}{k+2}}} \nonumber \\
&&
\end{eqnarray}
\end{center}
where $\rho = \left[ \omega^{(k+2)} (q_n^*)\right ]^{\frac{1}{k+2}}$. 
From the above, we can see that near a  $k$-th order front and for large times, the probability density $p(n,t)= |\psi(n,t)|^2$ scales as $\frac{1}{t^{2/k+2}}$ over a spatial region of extent $\approx t^{1/{k+2}}$.  Hence near ordinary fronts, $p(n,t) \approx \frac{1}{t}$ while it exhibits sub-diffusive behaviour (smaller than $1/t$) at extremal fronts. In particular, near the maximal fronts, the probability density scales as $1/t^{2/3}$~\citep{Toro,QW_Krapivsky2}.   
\begin{figure}[htp!]     
\subfigure{\label{fig:prob_dist_g_0}}
{\includegraphics[width=8.5cm,height=4.5cm,keepaspectratio]{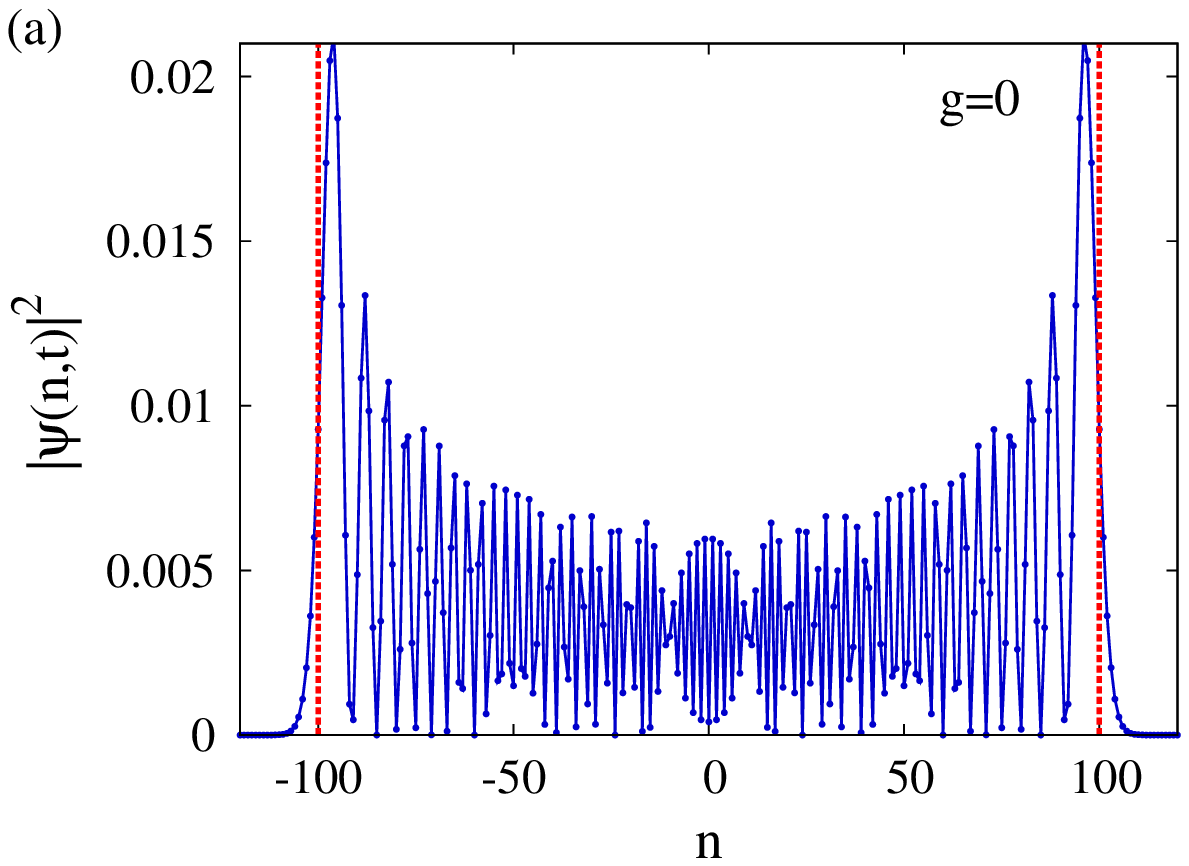}}

\subfigure{\label{fig:pd_g_1by8}}{\includegraphics[width=8.5cm,height=4.5cm,keepaspectratio]{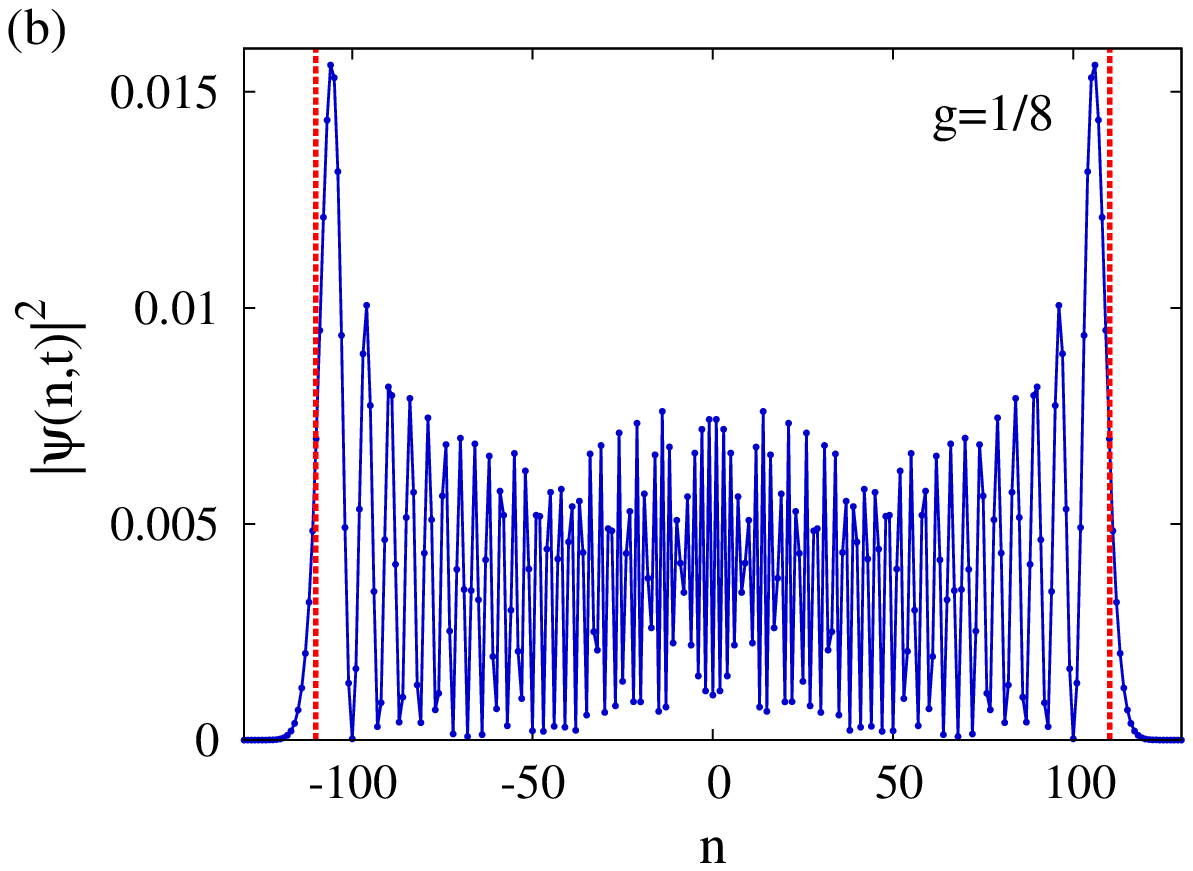}}

\subfigure{\label{fig:pd_g_1by4}}{\includegraphics[width=8.5cm,height=4.5cm,keepaspectratio]{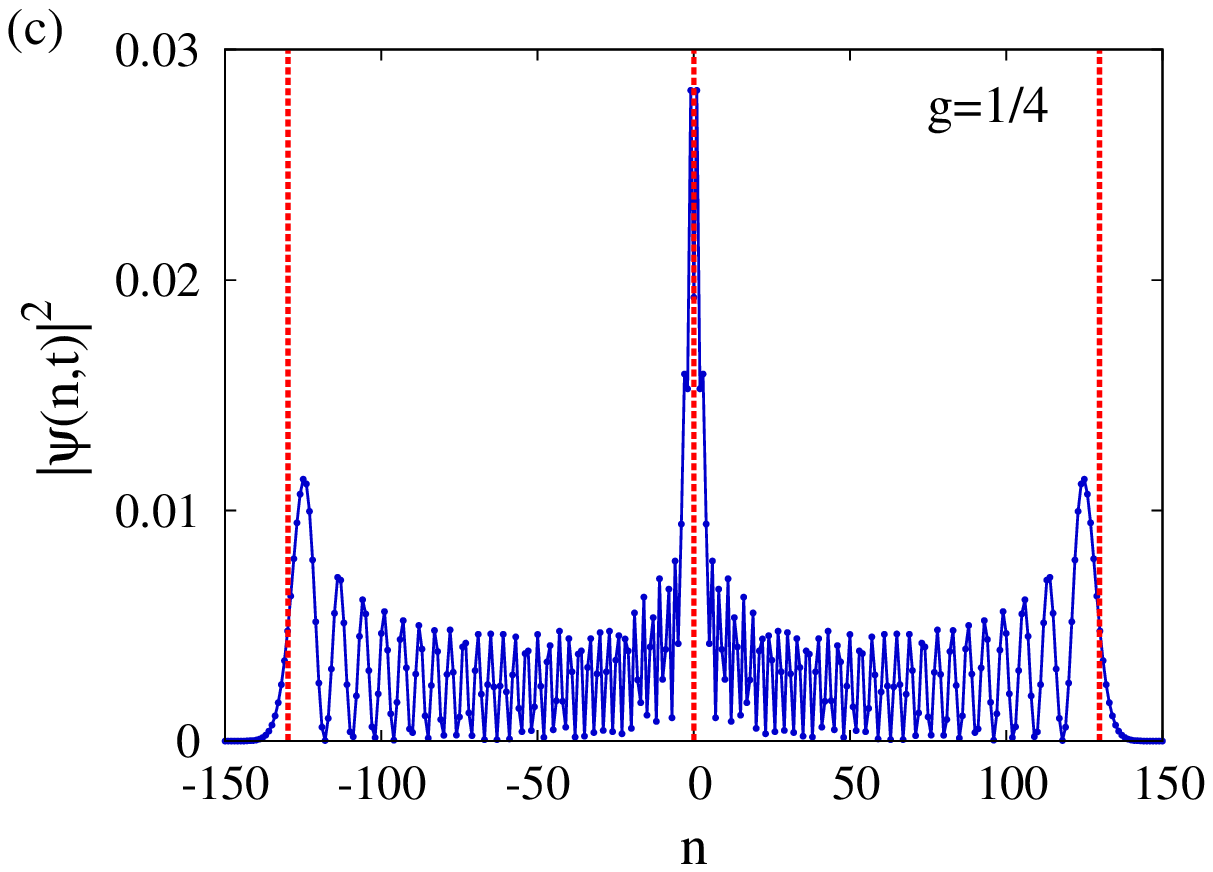}}

 \subfigure{\label{fig:pd_g_1by2}}{\includegraphics[width=8.5cm,height=4.5cm,keepaspectratio]{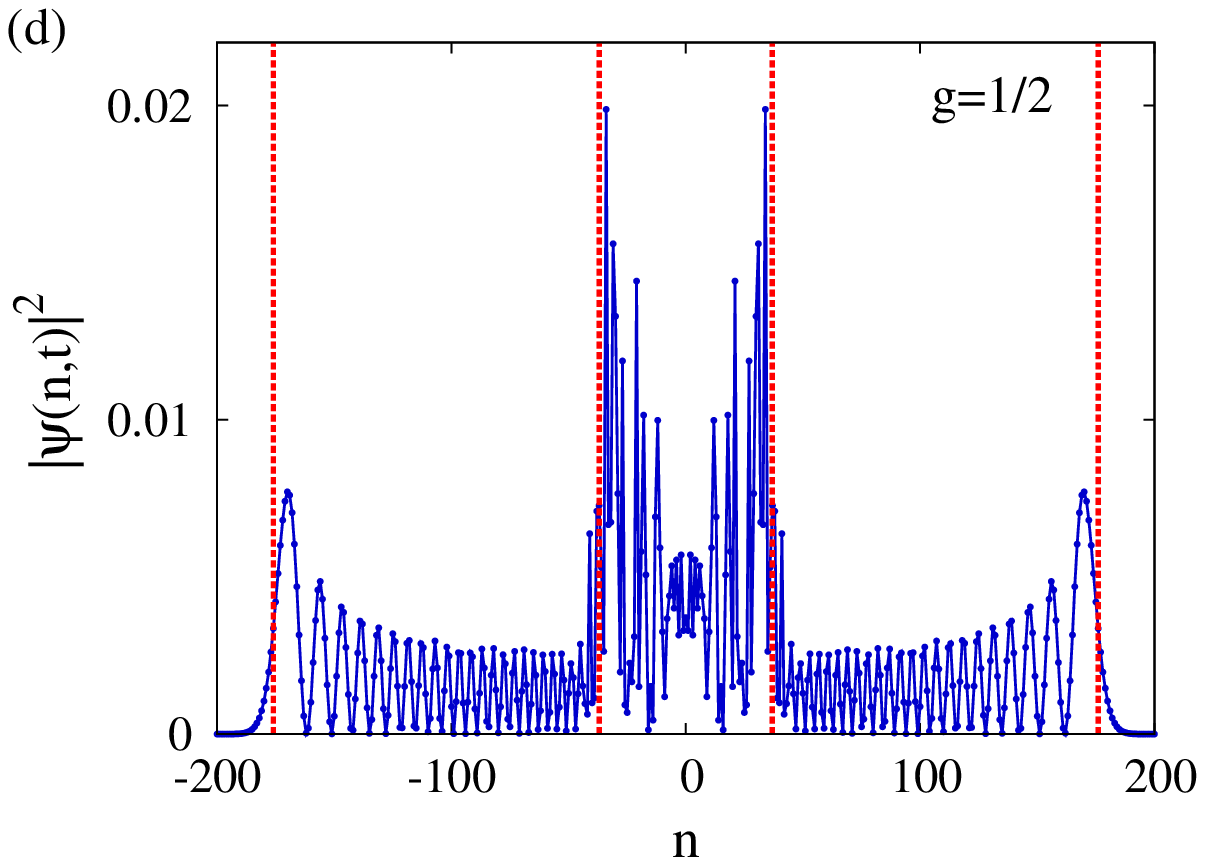}}
\caption{Local probability distribution profiles as a function of the lattice position $n$ (in units of the lattice spacing $a$) at time $t=50$ (in units of $1/g_1)$) for representative NNN hopping strengths $g$. The vertical red lines in the plots denote the theoretical positions of the extremal fronts. Additional internal extremal fronts appear when $g \geq g_c$.}\label{fig:prob_dist_g}      
\end{figure}

In general, from the saddle point approximation solutions, one  finds therefore, that there is an allowed region ($|n|<n_e$), within which the local probability density shows oscillatory behaviour and decays inversely with time  and a  forbidden region ($|n|>n_e$) lying outside the allowed region where the probability density decays exponentially~\citep{Toro, QW_Krapivsky2}. First order extremal fronts moving with the maximum of the group velocity $\pm v_e$ bound the allowed region. Additional extremal fronts can occur in the interior of the allowed region depending on the dispersion relation. There is an anomalous sub-diffusive scaling behaviour of the probability density near the extremal fronts with the exponent depending on the order of the front.

\section{\label{sec:global}{Global Scaling of the cumulative probability distribution}}

 We investigate in this section, the light-cone and front  dynamics by studying the  global scaling of the cumulative probability distribution using analytic and numerical methods. We also discuss the connections of the cumulative distribution profiles with the magnetization profiles in spin chain systems. 
We begin by defining the cumulative probability density $\Phi(n)$ as:
\begin{equation}\label{eq:cumulative_prob}
\Phi(n,t)\equiv‎‎\sum_{m\leq n}^{‎} \mid \psi(m,t)\mid ^2 =\sum_{m\leq n}^{‎} p(m,t)
\end{equation}
Conservation of probability implies that $ \Phi(-n, t)= 1-\Phi(n, t)$.

We restrict in the following and the rest of the paper to the case where only nearest and next-nearest neighbour hopping is allowed \citep{QW_Krapivsky2}. We also define the dimensionless ratio $g \equiv \frac{g_2}{g_1}$. From  Eq.~\ref{eq:dispersion} and Eq.~\ref{eq:front}, it can be seen that for all $g$, there are two first order extremal fronts, corresponding to the maximal fronts moving  with the maximal group velocities $\pm v_{e}$. $v_e$ depends on $g$ and is given as~\citep{QW_Krapivsky2}:
 \begin{equation}\label{eq:max_velocity}
v_{e} =\sqrt{\frac{-1+320g^2+2048g^4+(1+128g^2)^{3/2}}{128g^2} }   
\end{equation} 
 Additional extremal fronts appear for NNN hopping strengths $g$ greater than $g_c=1/4$~\citep{QW_Krapivsky2}.  When $g=g_c=1/4$, there are three extremal front solutions. Two of the solutions correspond to the maximal fronts travelling with maximal velocities $\pm v_e$. The third extremal front occurs at the wave vector value  $q=\pi$. This is a second order front since both the first and second derivative of $v(q)$ vanish at this $q$ value.  Also, the front has a  zero velocity $v_i=0$.  For  $g>g_{c}$,  there are two additional extremal fronts which move with velocities $\pm v_{i}$ given by \citep{QW_Krapivsky2}: 
 \begin{equation}\label{eq:int_velocity}
 v_{i} =\sqrt{\frac{-1+320g^2+2048g^4-(1+128g^2)^{3/2}}{128g^2} } 
 \end{equation}
The maximal velocities increase with $g$. The additional internal extremal front velocities which appear at $g>g_c$ also increase with $g$.

      The plots of the probability distribution profiles obtained from a numerical evaluation of the integral in Eq.~\ref{eq:fourier_integral} at a time $t=50$ are shown for representative values of $g$ in Fig.~\ref{fig:prob_dist_g}. Such plots have been shown earlier in Ref.\citep{QW_Krapivsky2}. We include the plots here for the sake of completeness. We can see from Fig.~\ref{fig:prob_dist_g} that for all $g$, the wave packet spreads with time with the propagation bounded by the maximal front velocity with a maximal spread of the wave packet given by $2 n_e = 2 v_{e} t$. The probability distribution is symmetric around the origin $n\leftrightarrow -n$ due to the reflection symmetry about the origin. The appearance of additional extremal fronts in the interior of the allowed region when $g \geq g_c$ can also be seen from Fig.~\ref{fig:prob_dist_g}. For $g=g_c$, we can see from Fig.~\ref{fig:pd_g_1by4} that there is an additional extremal front with velocity $v_i(q) =0$ at the origin while for $g>g_c$, two additional first order extremal fronts travelling with velocities $v_i$ given in Eq.~\ref{eq:int_velocity} occur at the sites $\pm n_i = v_i t$ as shown in Fig.~\ref{fig:pd_g_1by2}.  
      
\begin{figure}[htp]           
 
  \subfigure{\label{fig:velocity_cum_prob_a}}{\includegraphics[width=8cm,height=5cm,keepaspectratio]{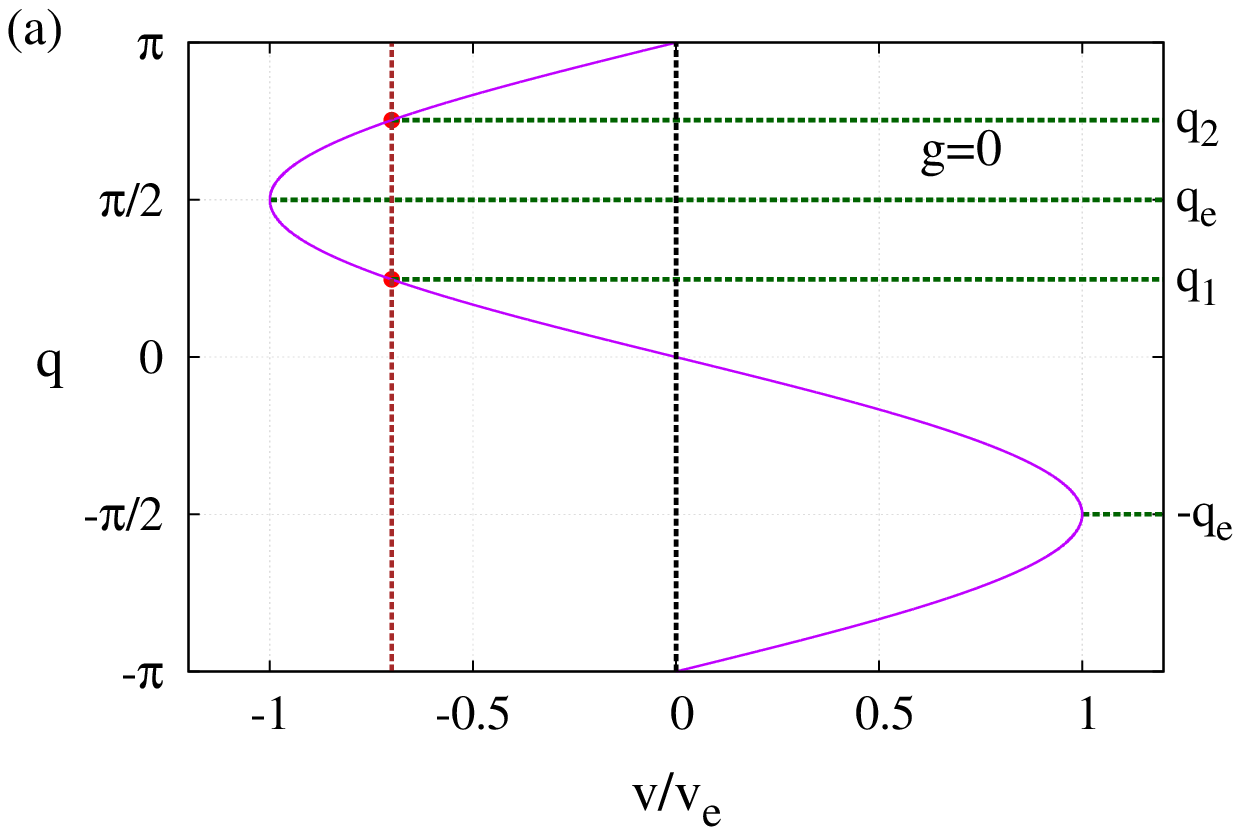}}
  
  \subfigure{\label{fig:velocity_cum_prob_b}}{\includegraphics[width=8cm,height=5cm,keepaspectratio]{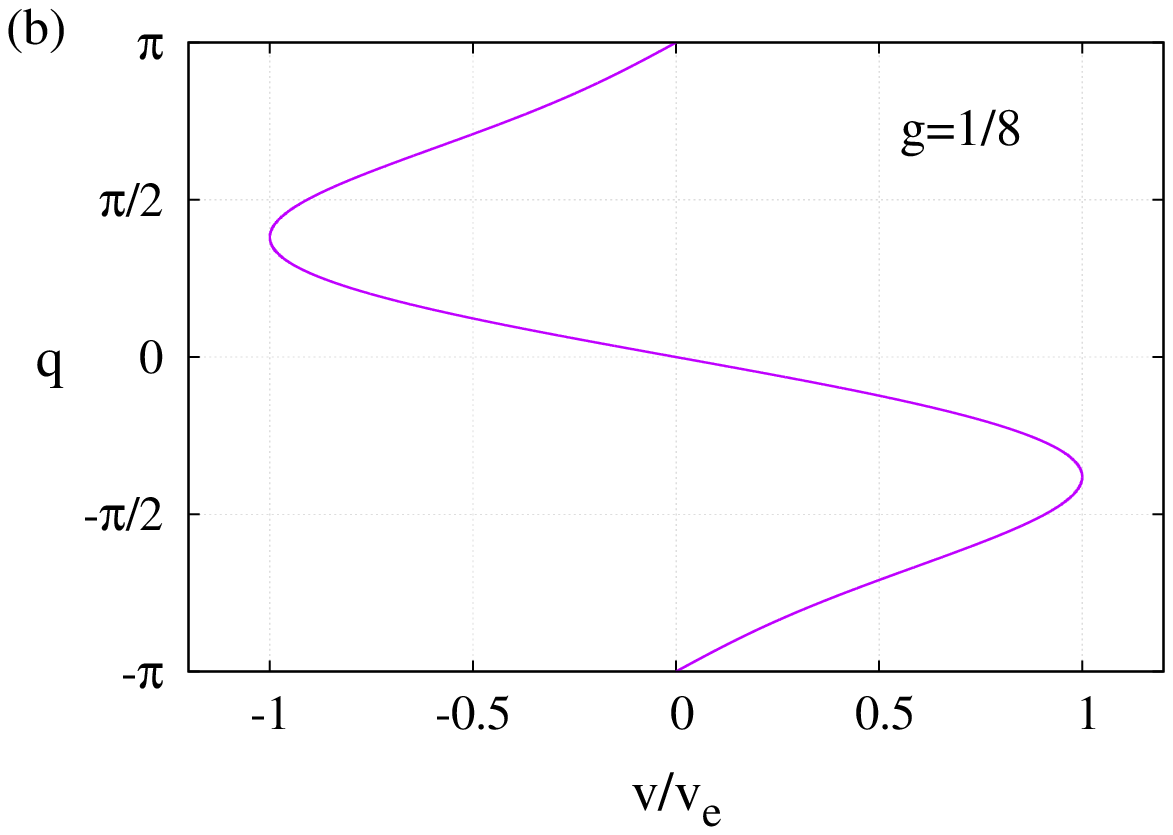}}
  
  \subfigure{\label{fig:velocity_cum_prob_c}}{\includegraphics[width=8cm,height=5cm,keepaspectratio]{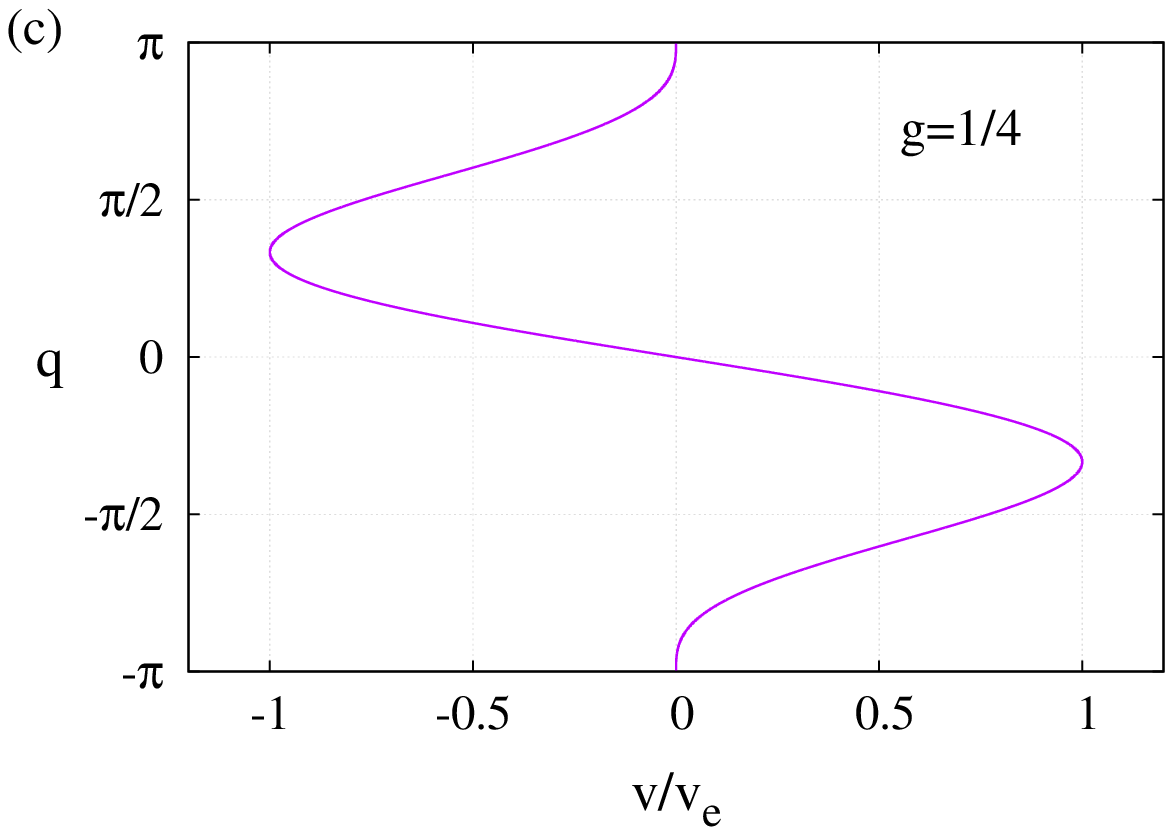}}
  
  \subfigure{\label{fig:velocity_cum_prob_d}}{\includegraphics[width=8cm,height=5cm,keepaspectratio]{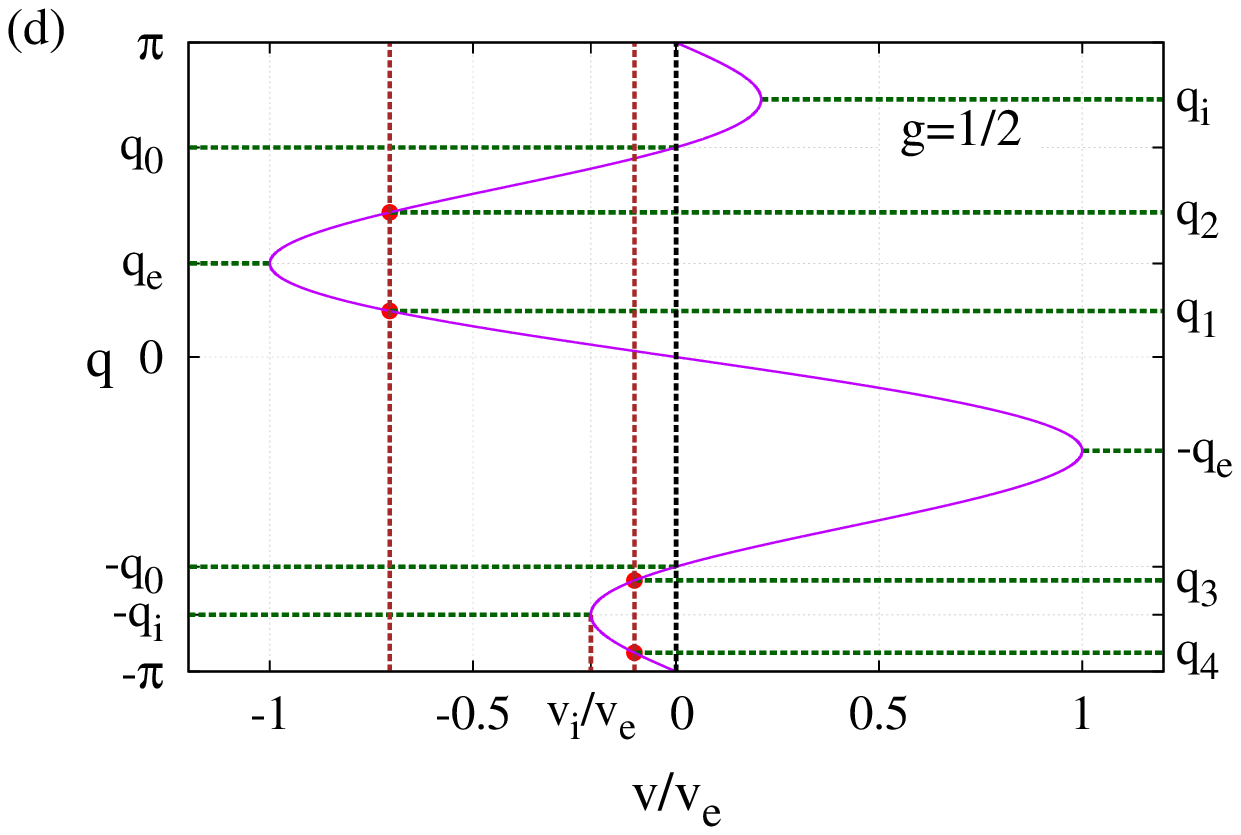}}
 \caption{ The dependence of the wave vector $q$ (in units of the inverse lattice spacing $1/a$) on the velocity $v$ (in units of the maximal velocity $v_e$) shown graphically for representative $g$ values.} \label{fig:velocity_cum_prob}    
\end{figure}
The causal structure is already apparent in the behaviour of the probability density. The  'allowed' region corresponds to a 'causal cone' inside which information is contained while the 'forbidden' region, inside which the probability density becomes exponentially small lies outside the causal cone.  We now show the existence of a global scaling form for the cumulative probability: $\Phi(n,t)=\Phi(\frac{n}{t})$. From the saddle point solutions $v(q)=\frac{n}{t}$  and from the velocity - wave vector dependences which we show graphically in Fig.~\ref{fig:velocity_cum_prob} for representative $g$ values,  we can see that the cumulative probability for the particle to be at site $n$ is given by the total  fraction of the density of excitations which arrive at site $n$ from the initial position $n=0$ with the velocity $v(q)> \frac{n}{t}$. 
Consider first the case  $g \leq g_c$.  Let us assume $n<0$. From the $q-v$ plots (Fig.~\ref{fig:velocity_cum_prob_a},   \ref{fig:velocity_cum_prob_b},  \ref{fig:velocity_cum_prob_c}), we can see that the fraction of density of excitations arriving at site $n$ with velocity $v(q) > n/t$ is given by $(q_2 -q_1)/2 \pi$. Here $q_1$ denotes the branch corresponding to $0>\frac{v}{v_e}>-1,  \quad  0 \leq  q \leq q_e$ and $q_2$ the branch corresponding to $ -1<\frac{v}{v_e}<0,   \quad q_e\leq  q \leq  \pi$. Also, $q_e= v^{-1}(n_e/t)$.  Note also that excitations on branch $1$ correspond to those with $\lambda (q) \equiv \omega^{(2)}(q)/v_e <0$ while excitations on branch $2$ correspond to those excitations with $\lambda (q) \equiv \omega^{(2)}(q)/v_e >0$.  The values of $q$'s can be obtained by solving the saddle point equation $v(q) = n/t$.  A similar analysis can be done when $n>0$. Thus, the cumulative probability takes the scaling form:
\begin{equation}\label{eq: cum_prob_scaling_gless}
 \Phi(n,t) = \Phi\left(\frac{n}{t}\right)= \left \{
  \begin{aligned}
    &0;&& n \leq-n_e  \\
    &N\left(\frac{n}{v_{e}t}\right) = \frac{(q_2-q_1)}{2 \pi} ; &&  \mid n \mid\leq n_e\\
    &1; &&  n \geq n_e
  \end{aligned} \right.
\end{equation}
When  $g=0, q_e = \pi/2$ and solving for $q$ using the saddle point solution,  $\Phi(n,t)$ has a simple global scaling form $\Phi(\frac{n}{t})$:
\begin{equation}\label{eq: cum_prob_scaling_g0}
 \Phi(n,t) = \Phi\left(\frac{n}{t}\right)= \left \{
  \begin{aligned}
    &0 && \text{for}\ n \leq-n_e  \\
    &\frac{1}{2}+\frac{1}{\pi} \sin^{-1}\left(\frac{n}{v_{e}t}\right)   && \text{for}\ \mid n \mid\leq n_e\\
    &1 && \text{for}\ n \geq n_e
  \end{aligned} \right.
\end{equation}

 We next turn to the case $g \neq 0,\,\, g >g_c$. Let us assume $n<0$. From the $q-v$ plot (Fig.~\ref{fig:velocity_cum_prob_d}), we see  that the fraction of density of excitations at site $n$ is again given by $(q_2 -q_1)/2 \pi$, where we have denoted $q_1$ as the branch corresponding to $0>\frac{v}{v_e}>-1, \quad  0 \leq  q \leq q_e$ and $q_2$ as the branch corresponding to $ -1<\frac{v}{v_e}<0,  \quad q_e\leq  q \leq  q_0$, where $q_e= v^{-1}(n_e/t)$ and $q_0= v^{-1}(0)$. However, there is an additional contribution to the density of excitations at site $n, n_i <n <0$  given by $(q_3-q_4)/2 \pi$ where we have denoted $q_3$ as the branch corresponding to $-\frac{v_i}{v_e}<\frac{v}{v_e}<0,  \quad  -q_i \geq  q \geq -q_o$ and $q_4$ as the branch corresponding to $ 0>\frac{v}{v_e} >-\frac{v_i}{v_e},  \quad -q_i\leq  q \leq  -\pi$. We note that here excitations on branch $1$ and $4$ correspond to those with $\lambda (q) \equiv \omega^{(2)}(q)/v_e <0$ while excitations on branch $2$ and $3$ correspond to those excitations with $\lambda (q) \equiv \omega^{(2)}(q)/v_e > 0$.  For $g >g_c$,  the cumulative probability can therefore be written as :
\begin{equation}\label{eq: cum_prob_scaling_ggreat}
 \Phi(n,t) = \Phi\left(\frac{n}{t}\right)= \left \{
  \begin{aligned}
   &0 && n \leq-n_e  \\
   &N_1\left(\frac{n}{v_{e}t}\right); && \!\!\!\!\!\!\!-n_e \leq n \leq -n_i \\
  & N_1\left(\frac{n}{v_{e}t}\right) + N_2\left(\frac{n}{v_{i}t}\right); &&   |n| \leq n_i \\
  &N_1\left(\frac{n}{v_{e}t}\right); &&  n_i \leq n \leq n_e\\
  &1;&& n \geq n_e \\
   \end{aligned} \right.
\end{equation}
where we have defined :
\begin{eqnarray}
N_1\left(\frac{n}{v_{e}t}\right)&=& \frac{(q_1 - q_2)}{2 \pi} ; \quad |n| < n_e \\
N_2 \left(\frac{n}{v_{i}t}\right)& =& \frac{(q_3 - q_4)}{2 \pi}; \quad |n| < n_i
\end{eqnarray}

\begin{figure}[htp]           
  
  \subfigure{\label{fig:cum_prob_a}}{\includegraphics[width=8.5cm,height=4.5cm,keepaspectratio]{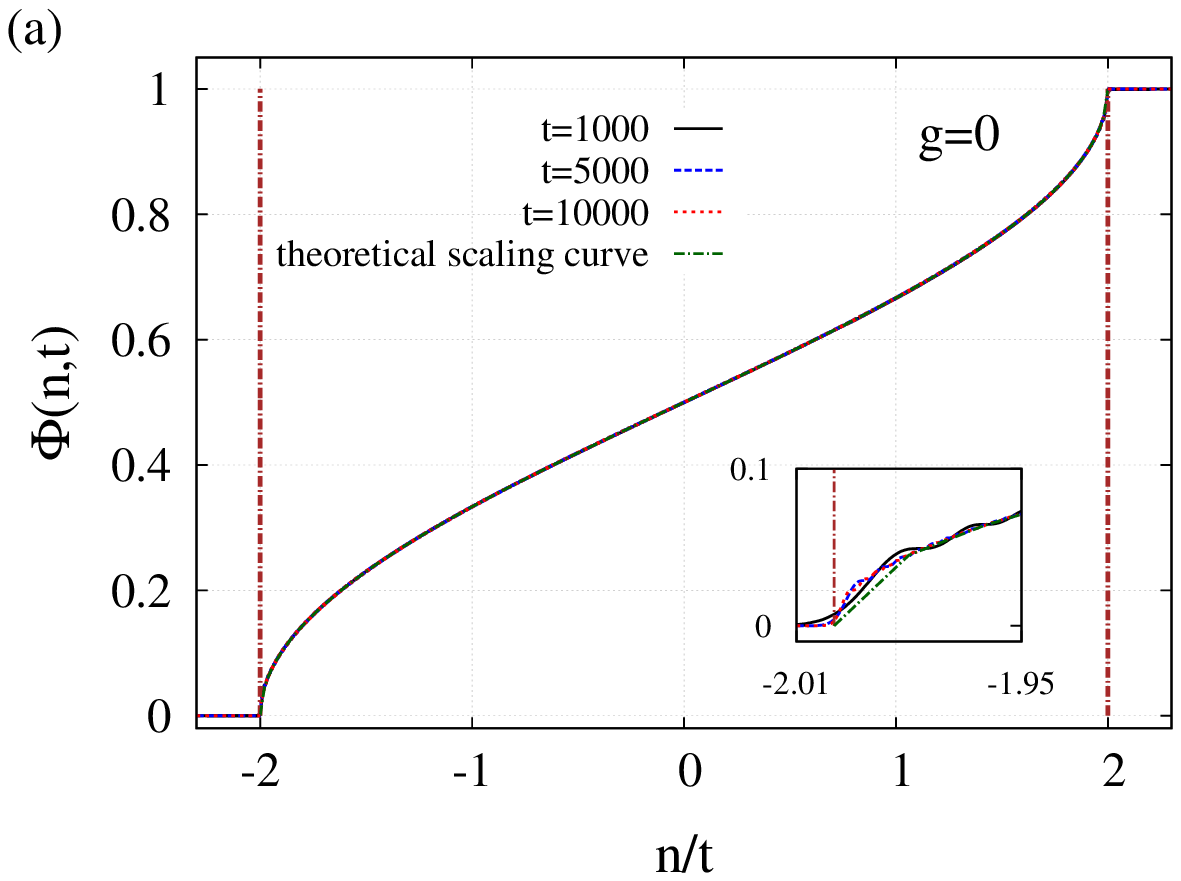}}
  
  \subfigure{\label{fig:cum_prob_b}}{\includegraphics[width=8.5cm,height=4.5cm,keepaspectratio]{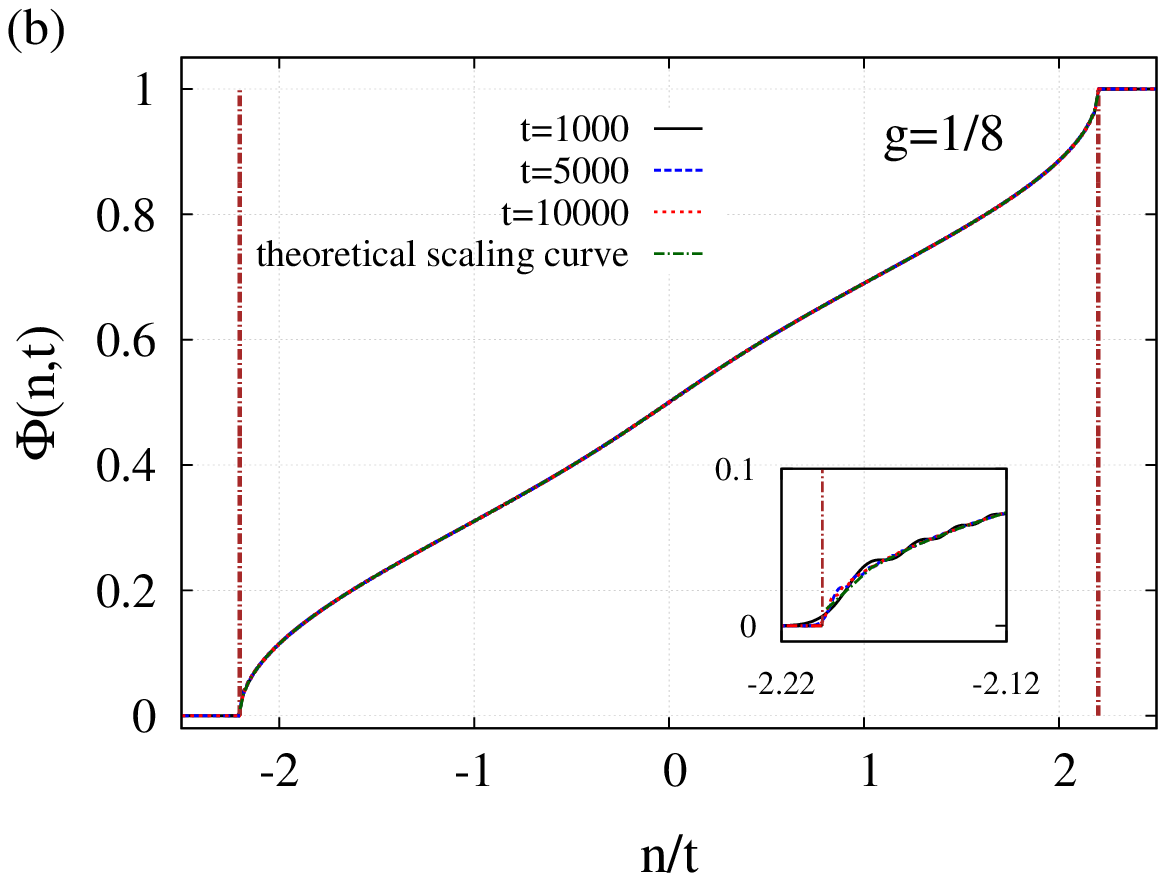}}
  
  \subfigure{\label{fig:cum_prob_c}}{\includegraphics[width=8.5cm,height=4.5cm,keepaspectratio]{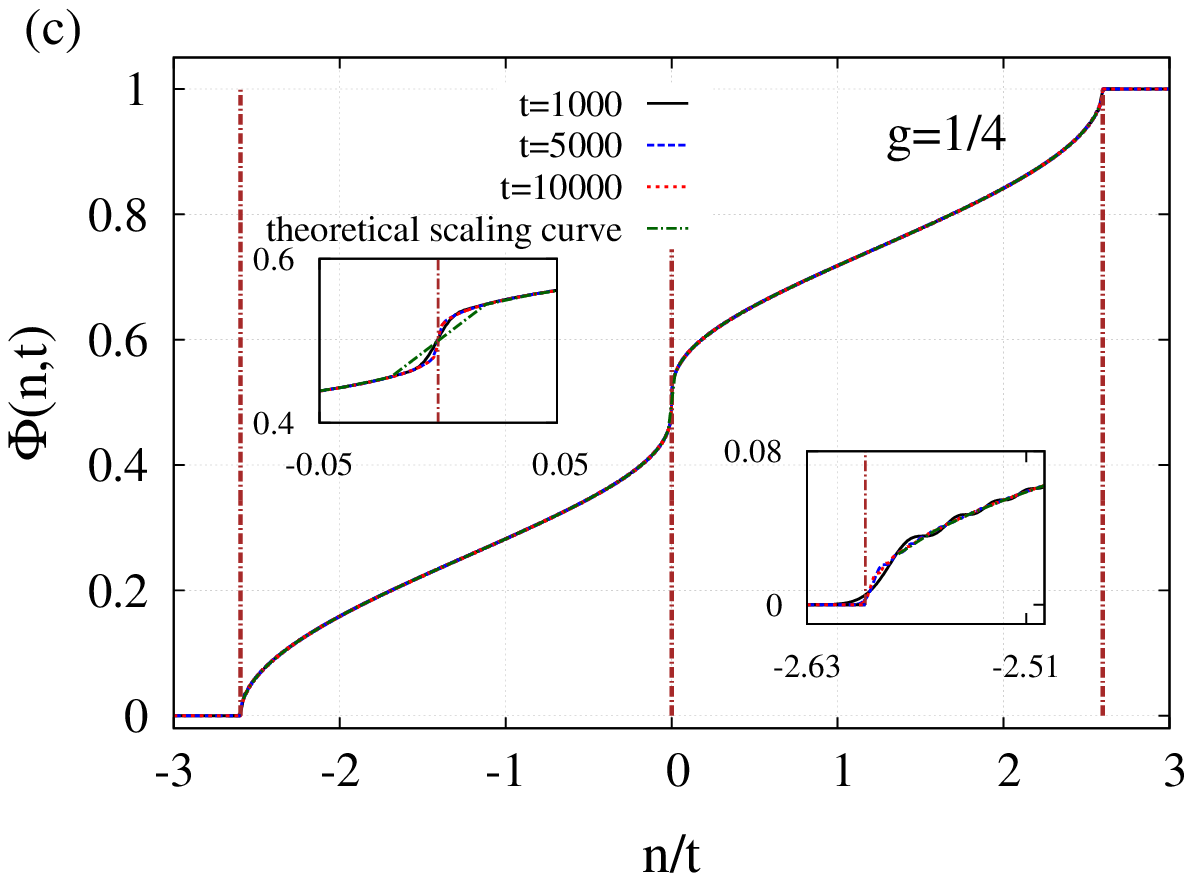}}
  
  \subfigure{\label{fig:cum_prob_d}}{\includegraphics[width=8.5cm,height=4.5cm,keepaspectratio]{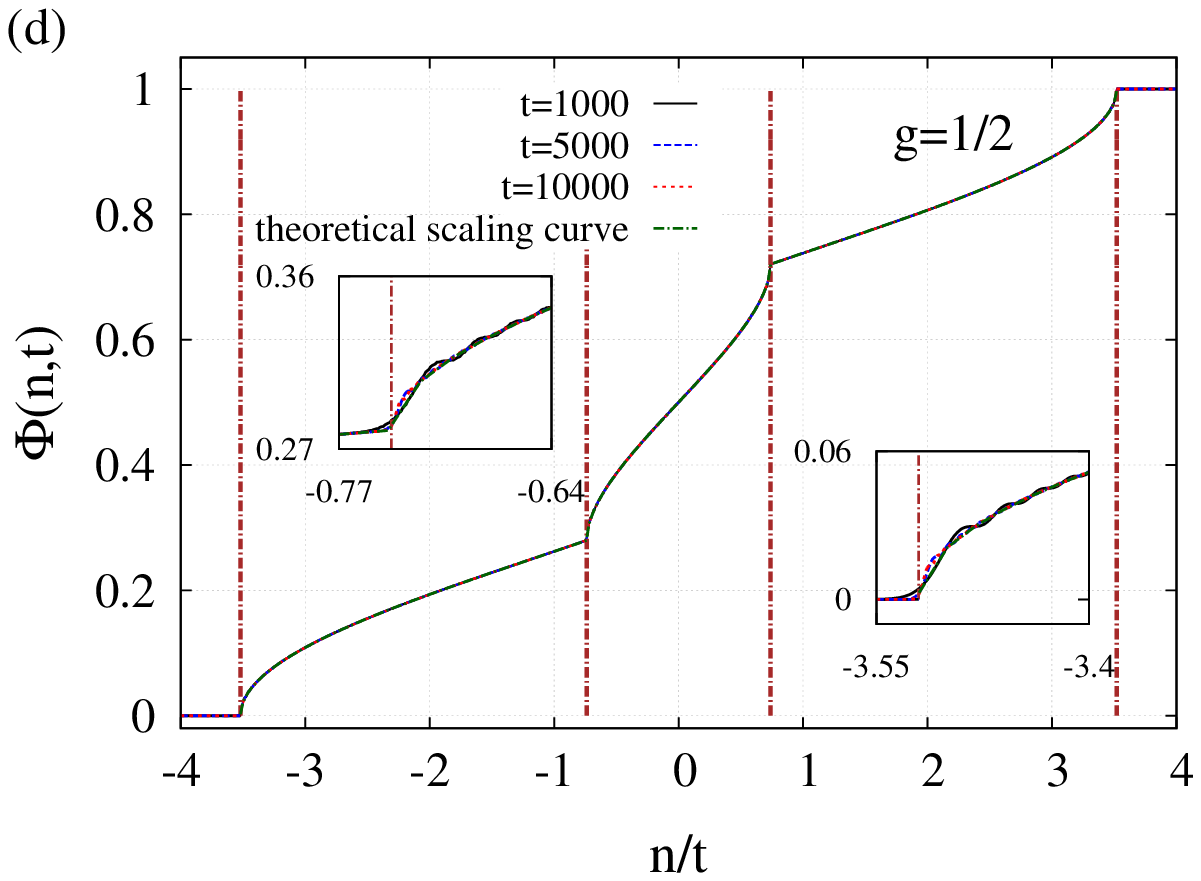}}
 \caption{The global scaling of the cumulative probability distribution function $\Phi(n,t)=\Phi(\frac{n}{t})$ as obtained from exact numerical calculations at different times $t=1000,5000,10000$ (in units of $1/g_1$) for representative NNN $g$ values as a function of velocity $n/t$ (in units of $ag_1$). The theoretical curve obtained using the saddle point solutions (Eq.~\ref{eq: cum_prob_scaling_gless} and Eq.~\ref{eq: cum_prob_scaling_ggreat})
is also shown for comparison.  The inset in each plot shows the deviation from global scaling  in the region near the fronts.} \label{fig:cum_prob}    
\end{figure}
The global scaling property $\Phi(n,t)\approx \Phi(\frac{n}{t})$ of the cumulative distribution functions  are demonstrated in Fig.~{\ref{fig:cum_prob}} for representative  $g$ values by using the  numerical solution for the local probability densities. The plot also shows the comparison with the scaling form obtained from the saddle-point solutions. From Fig.~\ref{fig:cum_prob}, we see that the numerical solutions for cumulative distribution functions match well with the  global scaling form obtained from the saddle point solutions (Eqs.~\ref{eq: cum_prob_scaling_gless}, \ref{eq: cum_prob_scaling_ggreat}). The distribution function is flat outside the 'causal cone' indicating that correlations exist only inside the 'causal cone'. The nature of the cumulative distribution profile inside the causal cone depends on the strength of $g$. For $g<g_{c}$ (Fig.~\ref{fig:cum_prob_b}), the behaviour is similar to that of the $g=0$ case (Fig.~\ref{fig:cum_prob_a}). There is an increase in the cumulative distribution near the origin though, for $g>0$.  For $g>g_{c}$, the emergence of additional  extremal fronts in the probability distribution can be seen from the cumulative distribution profile with the appearance of square root singularities at the internal extremal front locations as shown in Fig.~\ref{fig:cum_prob_d}. At $g=g_{c}$, there is a sharp finite discontinuity at the single internal front at the origin as can be seen in Fig. \ref{fig:cum_prob_c}.  

The global scaling form for the cumulative distribution function at large times can be attributed to the existence of a causal cone with ballistic propagation of ordinary fronts with a $1/t$ scaling. The two different kinds of global scaling when $g>g_c$ shows the emergence of an internal 'causal' cone.  It also indicates the existence of  two kinds of propagating excitations: one, bounded by a 'light' velocity, $v_e$ and the other bounded by a 'light' velocity $v_i$.
Near the extremal fronts,  there is a deviation from the global scaling as already seen from the saddle point solutions and shown in the inset in each of the sub figures in Fig.~\ref{fig:cum_prob}.  We observe here that the global scaling form of the cumulative distribution function for $g=0$ (Eq.~\ref{eq: cum_prob_scaling_g0}) and the cumulative distribution function obtained in Fig.~\ref{fig:cum_prob_a} correspond exactly to the magnetization profiles of time-evolved domain wall states in the Ising chain \citep{antal_pre59, sasvari_pre69}.  

\section{\label{sec:staircase}{Local Scaling and staircase structure}}

       We now study the local scaling structure near the extremal fronts. We show that the sub-diffusive scaling behaviour near extremal first order fronts gives rise to a staircase structure. The local scaling and staircase structure in the transition region near a maximal front can be investigated by studying the change in the cumulative probability from that at the maximal front located at, say, the right maximal front located at $n_{e}$ as:
\begin{equation}
\delta \Phi(n,t)\equiv \Phi(n_e,t)-\Phi(n,t)
\end{equation}
which can be expressed as:
\begin{subequations}\label{eq: deviation_cum_prob_general}
  \begin{empheq}[left={\delta \Phi(n,t)=\empheqlbrace\,}]{align}
      &0 && \text{for}\  n =n_e
        \label{eq: deviation_cum_prob_1_gen} \\
      & ‎\sum_{m=n}^{n_{e}‎}   p(m,t)  && \text{for}\ n < n_{e}
        \label{eq: deviation_cum_prob_2_gen}\\
      &- ‎‎\sum_{m=n_{e}}^{n‎}   p(m,t)   && \text{for}\ n > n_{e} ,
        \label{eq: deviation_cum_prob_3_gen}
    \end{empheq}
\end{subequations}

Consider first the case when there is only NN hopping  and where we have the exact solution for the local probability densities in terms of Bessel functions. In this case, the analysis described below is akin to that performed in Ref.\citep{sasvari_pre69} showing the emergence of an internal staircase structure in the magnetization profile of a time evolved domain wall initial state in the Ising chain. The change in the cumulative probability at the extremal maximal front located at $n_e$ can be written as:
\begin{subequations}\label{eq: deviation_cum_prob}
   \begin{empheq}[left={\delta \Phi(n,t)=\empheqlbrace\,}]{align}
      &0 && \text{for}\ n=n_e
        \label{eq: deviation_cum_prob_1} \\
      & ‎‎\sum_{m=n}^{n_e‎}  J_{m}^2(n_e) && \text{for}\ n < n_e 
        \label{eq: deviation_cum_prob_2}\\
      & -‎‎\sum_{m=n_e}^{n‎}  J_{m}^2(n_e)  && \text{for}\ n > n_{e} ,
        \label{eq: deviation_cum_prob_3}
    \end{empheq}
\end{subequations}

                                                                                                                                                                                                             Using the properties of Bessel functions in the asymptotic limit ($t\rightarrow \infty, n\rightarrow \infty$) ~\cite{abramowitz+stegun}, we can express the Bessel function in terms of scaled variables as:
\begin{equation}
J_{z_n}(n_e)\approx \frac{2^{1/3}}{\alpha^{1/3}(n_e)^{1/3}} Ai(\frac{-2^{1/3}z_n}{\alpha^{1/3}}) + \mathcal{O}((n_e)^{-1})
\end{equation}
where $Ai(z)$ is the Airy function and $z_n$ denotes the scaled distance from the front position: $z_n \equiv (n-n_e)/({n_e})^{1/3}$, $\alpha\equiv\frac{v^{(2)}(q_e)}{v_e}$. For asymptotically large times, the summation over $m$ in Eq.~(\ref{eq: deviation_cum_prob_2}) can be then converted to an integration over $z_m$ and the change in the cumulative probability obtained for $n<n_e$ as:
\begin{equation}
\delta \Phi(z_n,t)=- \frac{2^{2/3}}{{n_e}^{1/3} \alpha^{2/3}} \int_0 ^{z_n} F(z_m)dz_m \label{eq:airy_delta}
\end{equation}
where $F(z)=Ai^{2}\left(-2^{1/3} z/\alpha^{1/3}\right)$.
A similar analysis can be performed for  $n>n_e$. Thus, we find that that $\delta \Phi$ scales as $n_e^{1/3}$ near the front while distances to the front, $z$, scale as $n_e^{-1/3}$.  Also, from Eq.~\ref{eq:airy_delta}, we can see that the first and second derivative of $\delta \Phi$ vanish at the zeros of the Airy functions. The zeros of the Airy functions therefore correspond to stationary inflexion points of $\delta \Phi$. This leads to a staircase behaviour near the fronts. The existence of such steps shows the localized behaviour of the particle near the fronts, since the exponent of spatial spread, $(1/3)$, is smaller (sub-diffusive) than the generic spread $t^{1/2}$ given by Bessel functions. 

 We estimate the height, of each step by observing that in the front region, the cumulative probability (Eq.~\ref{eq: cum_prob_scaling_g0}) can be obtained as: 
\begin{eqnarray}
  \Phi(n,t)&=&\frac{1}{2}+\frac{1}{\pi} \sin^{-1}\left(1-\frac{z_n}{(n_e)^{2/3}}\right)
                 \approx \frac{1}{2}+\frac{1}{\pi}\left(\frac{\pi}{2}-\frac{\sqrt{2|z_n|}}{(n_e)^{1/3}}\right)\nonumber \\
               &\approx& 1-\frac{\sqrt{2|z_n|}}{\pi (n_e)^{1/3}}
  \end{eqnarray} 
Hence, defining $h_s={n_{e}}^{1/3}(\Phi(n,t)-\Phi(n_e,t))=n_e^{1/3}\delta \Phi(n_s,t)=\frac{\sqrt{2\mid z_{s}\mid}}{\pi}$, where the scaling variable $z_{s}$ corresponds to the  $s^{th}$ zero of $Ai(x)$. For asymptotic large s, $|z_{s}|\approx \frac{\alpha^{1/3}(3 \pi s)^{2/3}}{2}$ \,\citep{abramowitz+stegun} and the height is given as  $|h_{s}|= [3s\sqrt{\alpha}/{\pi^{2}}]^{1/3}$. The width of the steps in the staircase  can be obtained as the  distance between consecutive (non-stationary) inflexion points of $\delta \Phi$ which from Eq.~\ref{eq:airy_delta} are given by consecutive zeros of $Ai'(x)$. Thus the width can be expressed as:
$w_{s}=|\bar{z}_{s+1}- \bar{z}_{s}| \approx \left(\frac{\pi^{2}\alpha}{3s}\right)^{1/3} $\,\citep{abramowitz+stegun}. Here $\bar{z}_{s}$ and $\bar{z}_{s+1} $ denote consecutive zeros of $Ai'(x)$. The area below each step $h_s w_s = \alpha ^{1/2}$ is therefore a constant. We interpret this as quantization of probability of finding the particle in unit length or in other words, particle number quantization in unit length.
 
\paragraph*{} In the presence of NNN hopping, the local scaling and staircase structure is governed by the nature and number of  extremal fronts. Since the local scaling behaviour near an extremal front depends only on the order of the front, for the first order maximal fronts, the local scaling and staircase structure behaviour is similar to that for the $g=0$ case. We can obtain the wavefunction at the site $n\approx n_e$ by evaluating the Fourier integral (Eq.~\ref{eq:fourier_integral}) using the saddle point approximation. Expanding $\tilde \varphi(n,q) = \varphi(n,q)/v_e$ for $n$ near $n_e$, we can write:
\begin{eqnarray}
\tilde \varphi (n,q) &&\approx (q_e - \tilde \omega(q_e) ) +  \frac{n-n_e}{n_e}q_e + \frac{n-n_e}{n_e}(q-q_e) \nonumber \\
 && \qquad  - \frac{(q-q_e)^3}{3!} \tilde v^{(2})(q_e)
\end{eqnarray}
where $\tilde v^{(2)}(q_e) \equiv \frac{v^{(2)}(q_e)}{v_e}$. 
Then the wavefunction at site $n\approx n_e$ can be expressed as:
\begin{eqnarray}\label{eq:extremal_front_solution}
&& \psi(n,t) \nonumber \\
&&= \int_{-\pi}^{\pi} \frac{dq}{2 \pi}  ‎e^{i n_e  \tilde \varphi(n,q) }\nonumber \\
&& \approx \int_{-\pi}^{\pi} \frac{dq}{2 \pi}  ‎e^{i n_e [(q_e - \tilde \omega(q_e) ) +  \frac{n-n_e}{n_e}q_e + \frac{n-n_e}{n_e}(q-q_e) - \frac{(q-q_e)^3}{3!} \tilde v^{(2)}(q_e)]} \nonumber \\
& & = ‎e^{i n_e [(q_e - \tilde \omega(q_e) ) +  \frac{n-n_e}{n_e}q_e]} \int_{-\pi}^{\pi} \frac{ d q}{2 \pi} e^{i n_e \left[\frac{n-n_e}{n_e}(q-q_e) - \frac{(q-q_e)^3}{3} \frac{\alpha}{2}\right] } \nonumber \\
&& \nonumber \\
&&
\end{eqnarray}
where $\alpha \equiv \tilde v^{(2)}(q_e)$. Transforming to the scaled variables $z = \frac{n-n_e}{n_e^{1/3}}$  and 
$ \xi = n_e^{1/3} 2 ^{-1/3} \alpha^{1/3} (q-q_e)$, we can express the integral in the last line of Eq.~\ref{eq:extremal_front_solution} as: 
\begin{eqnarray}
&& \int _{-\pi}^{\pi} \frac{ dq }{ 2 \pi} e^{ i n_e \left[\frac{n-n_e}{n_e}(q-q_e) - \frac{(q-q_e)^3}{3} \frac{\alpha}{2} \right]} \nonumber \\
&&= \frac{2^{1/3}}{\alpha^{1/3}}\frac{1}{n_e^{1/3}}\int_{-\infty}^{\infty} \frac{d\xi}{ 2 \pi} e^{ i \frac{2 ^{1/3}}{ \alpha ^{1/3}}  z \xi} e^{- i \frac{ \xi^3}{3}}\nonumber \\
& &=\frac{2^{1/3}}{\alpha^{1/3}}\frac{1}{n_e^{1/3}} Ai\left(-\frac{2^{1/3}}{\alpha^{1/3}} z\right)
\end{eqnarray}
We can thus evaluate  $\delta \Phi= \sum_{m=n}^{n_e} p(m,t) $ in the transition region $n \approx n_e, n<n_e$ as ,  
\begin{equation}\label{eq:g_extremal_front_scaling}
\delta \Phi= \sum_{m=n}^{n_e} p(m,t) =- \frac{ 2^{2/3}}{\alpha^{2/3}}\frac{1}{n_e^{1/3}} \int _0^{z_n} Ai^2 \left(-\frac{2^{1/3}}{\alpha^{1/3}} z\right)d z
\end{equation}
From Eq.~\ref{eq:g_extremal_front_scaling}, we can see that we obtain local scaling and staircase behaviour near the maximal fronts  similar to that for the $g=0$ case. The width and height of the steps can be obtained as earlier from the inflexion points of the Airy functions and zeros of its derivative. Since $v_e$ and $\alpha$ depend on $g$, the width, height and the number of steps depend on $g$. However, it can be easily seen that the area under each step remains a constant as for the $g=0$ case. When $g>g_c$, the presence of the two additional internal first order extremal fronts lead to an additional internal staircase structure near the internal fronts at $n_i$ which again has a similar Airy scaling and staircase behaviour as that near the maximal fronts. However, the scaling is now with respect to the extremal velocity $v_i$ and front position $n_i$.

\begin{figure*}[htp]           
  \centering
  \subfigure{\label{fig:stair_cum_prob_a}}{\includegraphics[width=9cm,height=5.5cm,keepaspectratio]{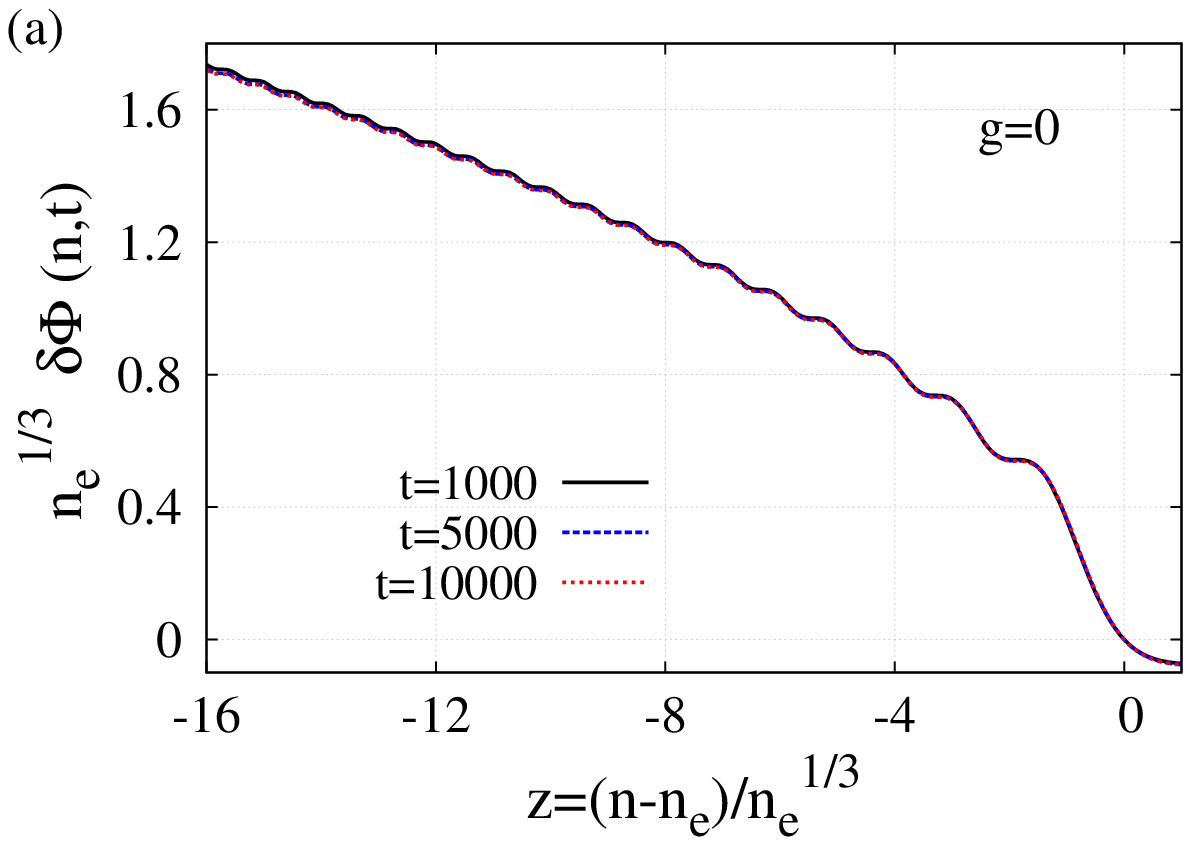}}\qquad 
\subfigure{\label{fig:stair_cum_prob_b}}{\includegraphics[width=9cm,height=5.5cm,keepaspectratio]{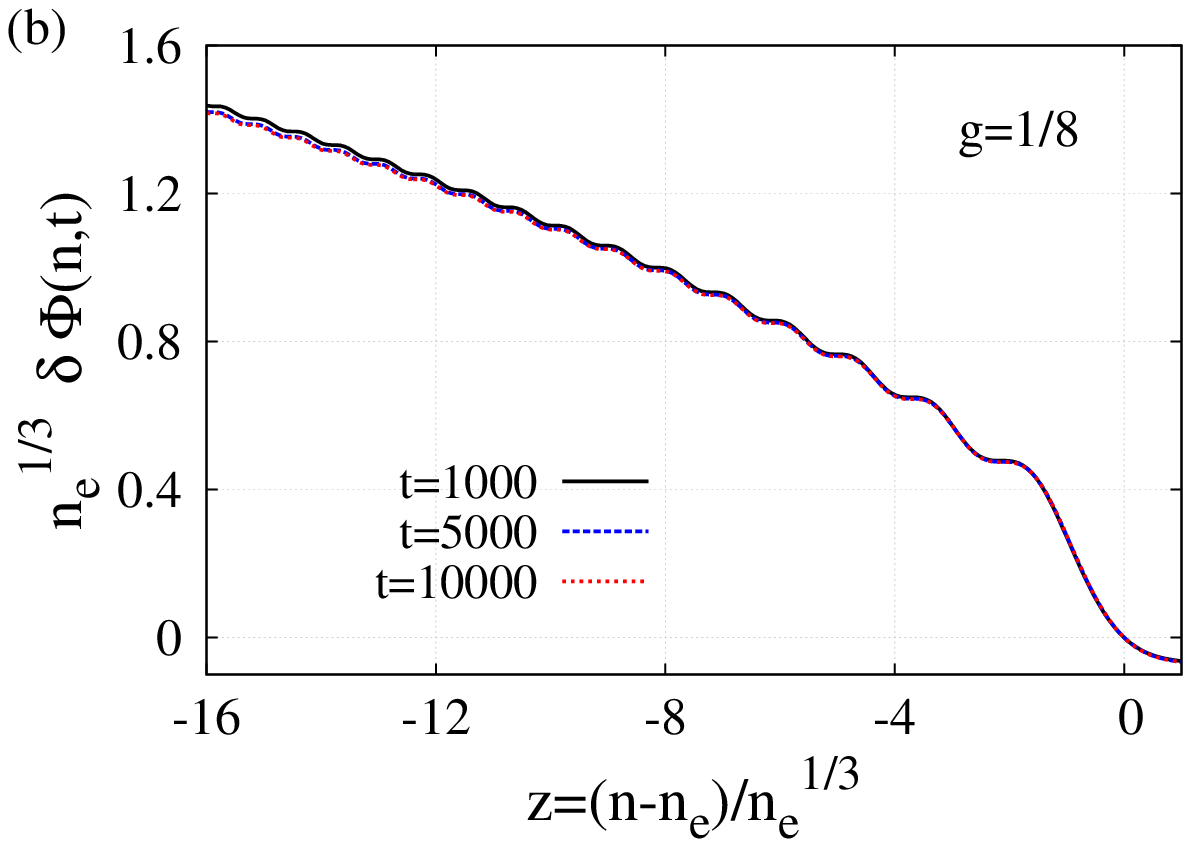}}\quad

\subfigure{\label{fig:stair_cum_prob_c}}{\includegraphics[width=9cm,height=5.5cm,keepaspectratio]{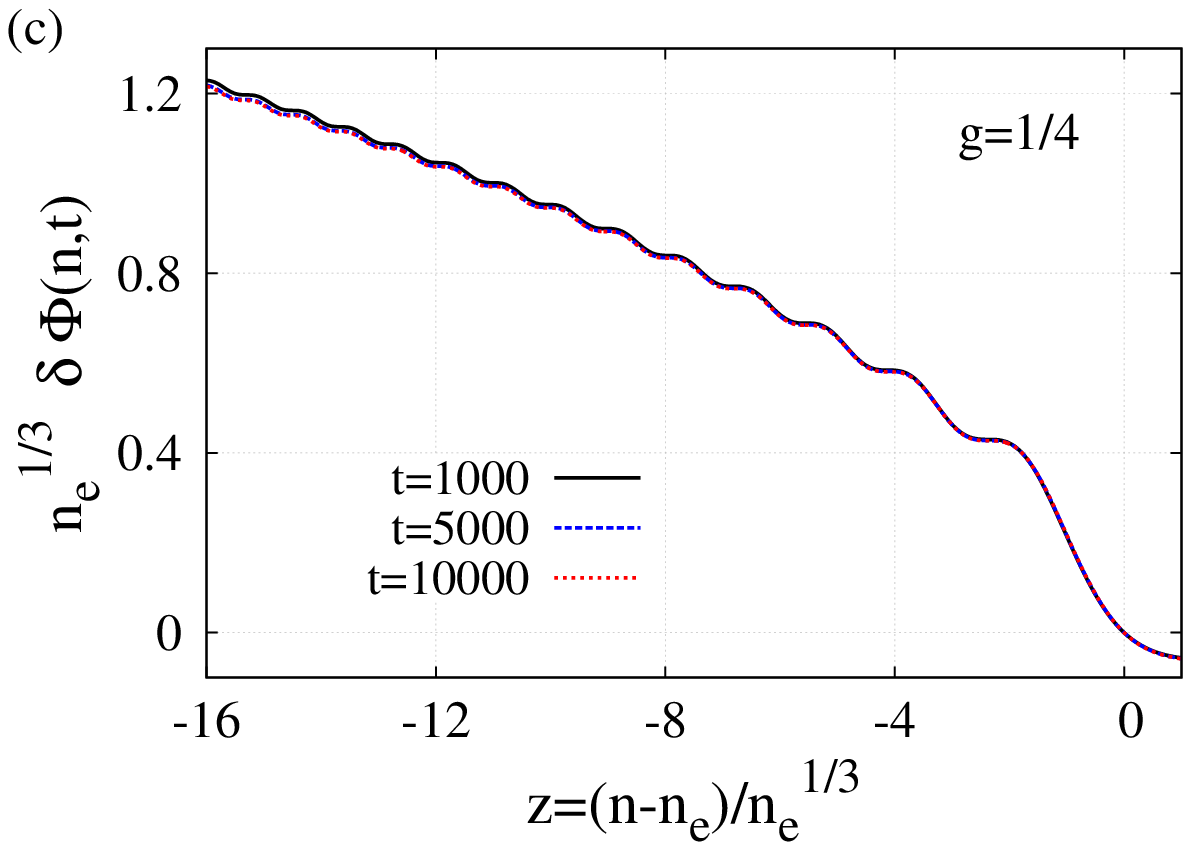}}\qquad
  \subfigure{\label{fig:stair_cum_prob_d}}{\includegraphics[width=9cm,height=5.5cm,keepaspectratio]{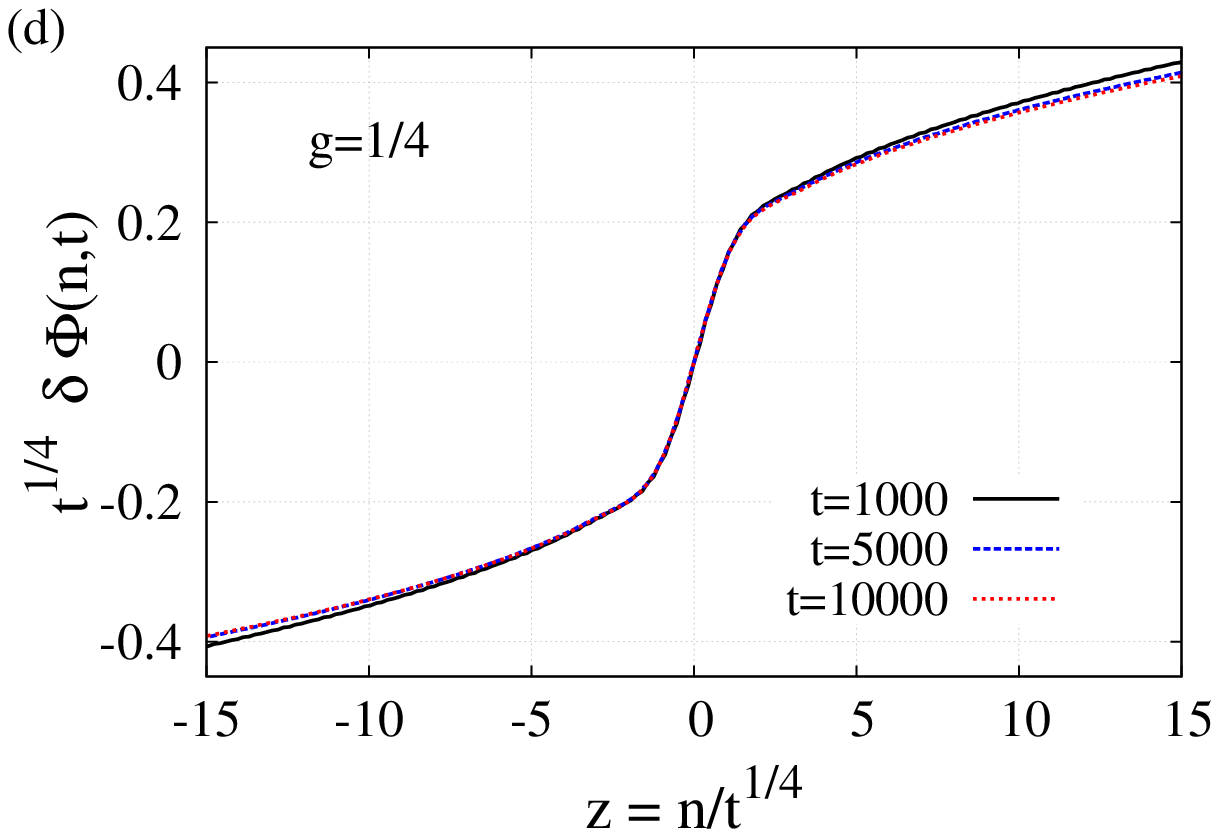}}\quad

  \subfigure{\label{fig:stair_cum_prob_e}}{\includegraphics[width=9cm,height=5.5cm,keepaspectratio]{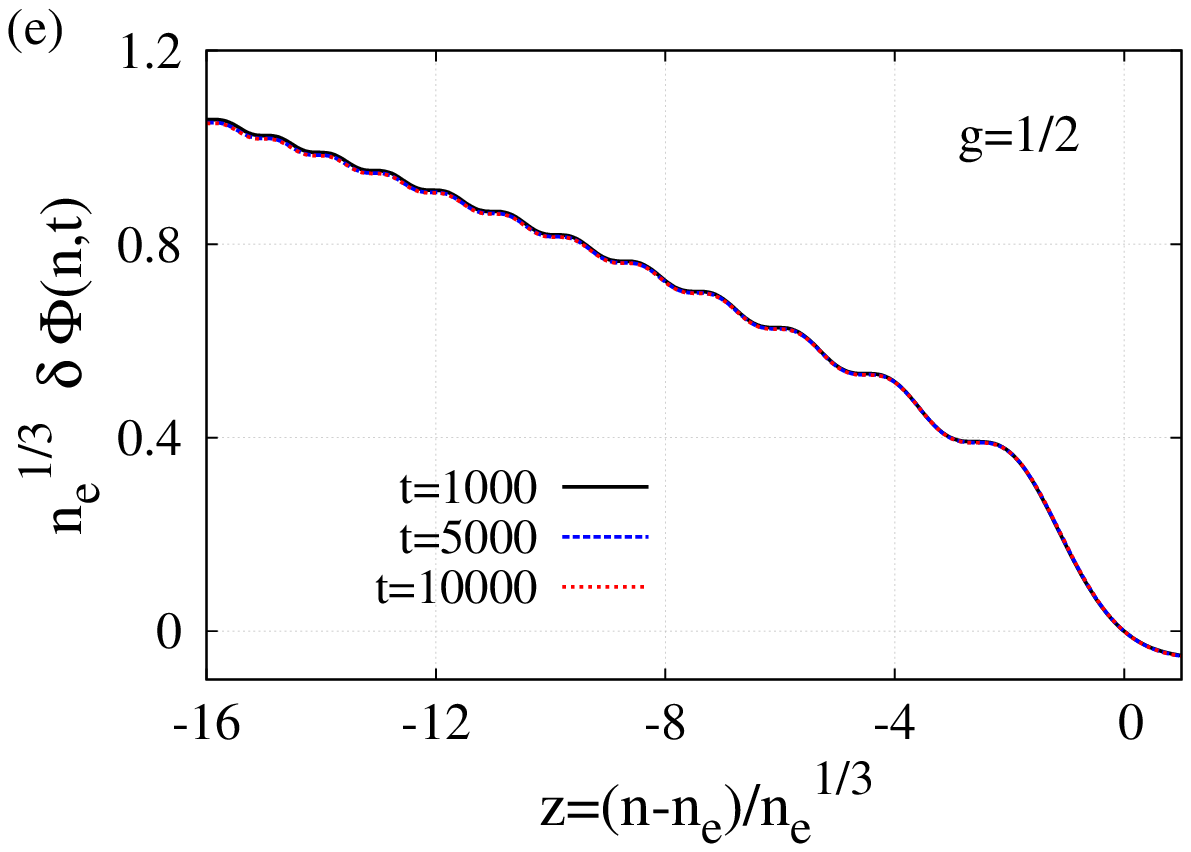}}\qquad
  \subfigure{\label{fig:stair_cum_prob_f}}{\includegraphics[width=9cm,height=5.5cm,keepaspectratio]{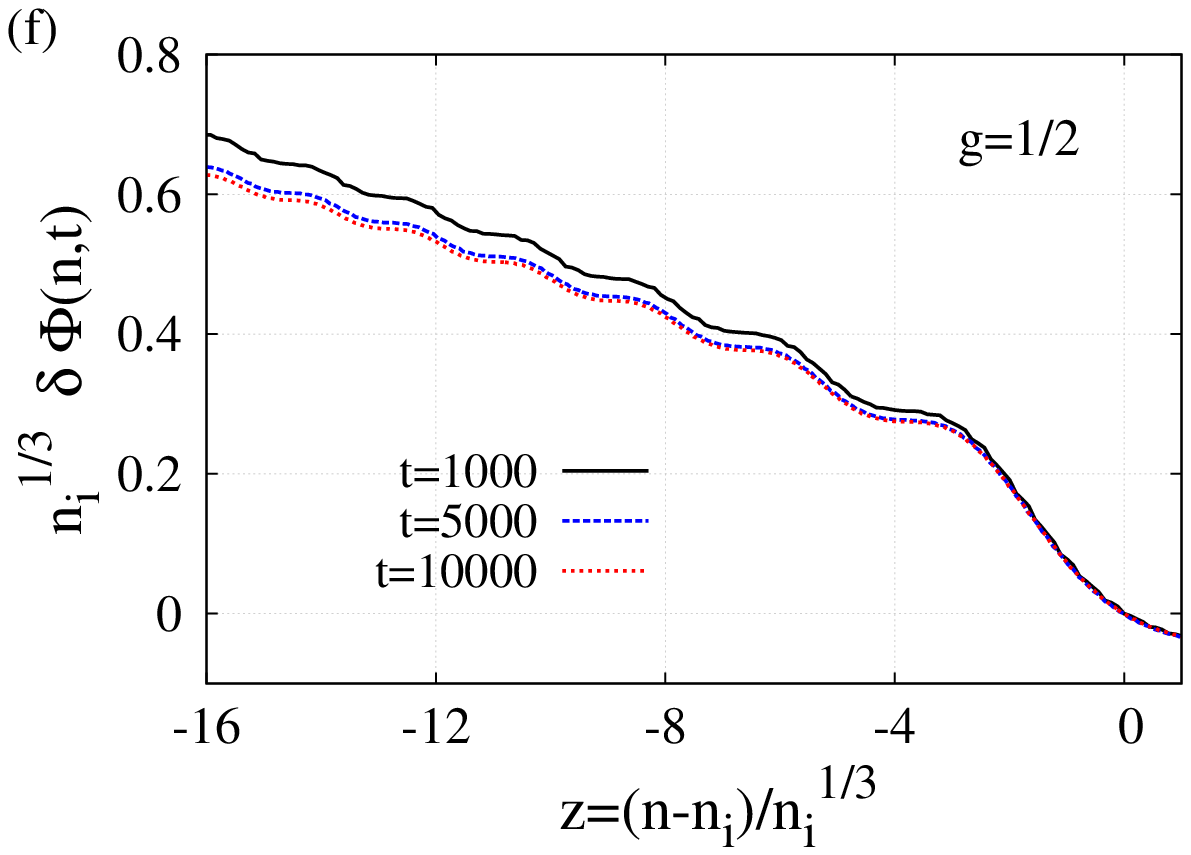}} \quad
\caption{ The local scaling and internal staircase structure near the extremal fronts as obtained from exact numerical calculations for representative NNN $g$ strengths and at different times $t=1000, 5000, 10000$ (in units of $1/g_1$). Panels (a,b,c,e) show the staircase structure near the right maximal front at $n_e$ for representative $g$ values. For any $g$, there is an Airy scaling near the maximal front with distances from the front scaling as $z=(n-n_e)/{n_e}^{1/3}$ while the cumulative probability measured from $\Phi(n_e,t)$ scales as $n_e^{1/3}\delta \Phi$. The scaled heights and widths obtained from the numerics match reasonably well with the (scaled) zeros of the Airy function and its derivative as predicted by the theoretical saddle point analysis for the different $g$ values. Similar local scaling and staircase structure is found at the left maximal front. Panel (d) shows the local second order extremal front scaling near the 
origin at the critical NNN hopping strength $g=g_c$. In this case, the distance from the internal front at the origin is measured in terms of the scaled variable $n/t^{1/4}$ while the cumulative probability measured from the value at the front $\Phi(n_i=0,t)$ scales as $t^{1/4}\delta \Phi$. We do not observe any internal staircase structure near the front. The numerics are in agreement with the theoretical analysis. Panel (f) shows the local scaling near the internal first order extremal front for the case $g=1/2$ at $n_i$ with the distance from the front $z$ scaling as $(n-n_i)/n_i^{1/3}$ and the cumulative probability measured from the front at $n_i$ scaling as $n_i^{1/3}\delta \Phi$. }\label{fig:stair_cum_prob}
\end{figure*}

For $g=g_c$, the additional extremal front occurring at the origin is a second order front. By a similar analysis as above, the wavefunction at a site $n\approx n_i =0$ can be written as:
\begin{equation}
\psi(n,t)= P \int_{-\pi}^{\pi} \frac{dq}{2 \pi} e^{ i t \left[ n \frac{(q-q_i)}{t} - \frac{(q-q_i)^4}{4 !} \omega ^{(4)}(q_i)\right]}
\end{equation}
where $q_i=v^{-1}(0)$. Here $P$ is a phase factor given as $P\equiv e^{-i(w(q_{i})t-{n q_{i}})}$. Defining $\beta \equiv \frac{\omega^{(4)}(q_i)}{3!}$ and transforming to the scaled variables $z = \frac{n}{t^{1/4}}$ and 
$ \xi = t^{1/4} \beta^{1/4} (q-q_i)$, we can write: 
\begin{equation}
\psi(n,t)= \frac{P}{\beta^{1/4}t^{1/4}}\int_{-\infty}^{\infty} \frac{d\xi}{2 \pi} e^{ i \frac{z \xi}{\beta^{1/4}}} e^{- i \frac{\xi ^4}{4}}
\end{equation}
The change in the cumulative probability $\delta \Phi= \sum_{m=n}^{n_i=0} p(m,t) $ in the transition region near $n=n_i=0$ can be then evaluated as:
\begin{equation}
\delta \Phi(z_n) = -\frac{1}{\beta^{1/2} t^{1/4}}\int _0^{z_n} dz\,\, I^2 (z_m)
\end{equation}
where we have defined $I(z) =\int_{-\infty}^{\infty} \frac{d\xi}{2 \pi} e^{ i \frac{z \xi}{\beta^{1/4}} }e^{- i \frac{\xi ^4}{4}}$.
Hence the stationary inflexion points of $\delta \Phi'(z)$ should correspond to the zeros of $I(z)$. We do not have a closed expression for $I(z)$, but we conjecture that $I(z)$ has no real zeros. We do not therefore expect any staircase structure near the front.From the behaviour of $I(z)$ for small $z$, we find that there is only a single sharp step of zero width near the origin with an anomalous $t^{1/4}$ local scaling behaviour near $n=0$ given as:
\begin{equation}
 t^{1/4}\delta \Phi(n,t)\approx \left \{
  \begin{aligned}
    &-C \dfrac{\mid n \mid} {t ^{1/4}} && \text{for}\ n < 0  \\
    &C\dfrac{\mid n \mid }{t ^{1/4}}   && \text{for}\  n>0\\
  \end{aligned} \right.
\end{equation} 
where $C$ is a constant.
We note that it was pointed out in Ref.\citep{QW_Krapivsky2} that the local probability shows an anomalous $1/t^{1/2}$ scaling near the internal front at the origin for $g=g_c$. The lack of any internal staircase structure near the front points to the absence of any propagating excitation and hence, localization.

    We show in Fig.~\ref{fig:stair_cum_prob}, the local scaling and staircase structure near the extremal fronts as obtained from the numerical calculations for different times and for representative $g$ values. We also compare with the predictions from the saddle point analysis discussed above. We plot the local scaling behaviour near the right maximal front at $n=n_e$ for the different $g$ values in Figs.~\ref{fig:stair_cum_prob_a}, ~\ref{fig:stair_cum_prob_b}, ~\ref{fig:stair_cum_prob_c}, ~\ref{fig:stair_cum_prob_e}. It can be seen from the plots that the maximal front region is characterized by a staircase structure having the Airy scaling behaviour given by Eq.~(\ref{eq:g_extremal_front_scaling}). We also find that the scaled heights and widths obtained from the numerical calculations match reasonably well with the (scaled) zeros of the Airy function and its derivative as predicted by the theoretical saddle point analysis for the different $g$ values. The (scaled) area under each step remains unity as shown in  Table~\ref{tab:Areas} and Table~\ref{tab:Areas_2} (left column) for the different $g$ values.   

\begin{table}
\begin{ruledtabular}

\begin{tabular}{ccccccccc}
&\multicolumn{6}{c}{$\text{Area below the steps}$}\\
Step&$g=0$&$g=1/8$&$g=1/4$
&$g=0$&$g=1/8$&$g=1/4$\\ 
&\multicolumn{3}{c}{$\text{t=5000}$}&\multicolumn{3}{c}{$\text{t=10000}$}\\
\hline 
& & & & & & &  \\

 1&0.9522&0.9611&0.9530&0.9531&0.9525&0.9525 \\
 2&0.9148 &0.9150&0.9113&0.9130&0.9129&0.9118\\
 3&0.9228&0.9206&0.9273&0.9196&0.9182&0.9187\\
 4&0.9239&0.9230&0.9283&0.9257&0.9252&0.9250\\
 5&0.9322&0.9338&0.9223&0.9313&0.9300&0.9287\\
 \end{tabular}
\end{ruledtabular}

\caption{\label{tab:Areas}The (scaled) area (in units of lattice constant) below the steps in the local staircase structure obtained from the numerical results for the cumulative distribution function at times $t=5000,10000$ (in units of $1/g_1$) (Figs.~\ref{fig:stair_cum_prob_a},~\ref{fig:stair_cum_prob_b},~\ref{fig:stair_cum_prob_c}) for different strengths $g$, $g\leq g_{c}$. It can be seen from the table that the (scaled) area below each step is independent of $g$ and remains nearly a unit constant.  
}
\end{table}  

\begin{table}
\begin{ruledtabular}
\begin{tabular}{ccccccccc}
&\multicolumn{5}{c}{$\text{Area below the steps (g=1/2)}$}\\
&\multicolumn{2}{c}{$\text{External front}$}&\multicolumn{2}{c}{$\text{Internal front}$}\\
 Step& \text{$t=5000$}& \text{$t=10000$}
&\text{$t=5000$} & \text{$t=10000$}  \\ \hline
\\
 1&0.9510&0.9513&1.0211&1.0035 \\
 2&0.9122&0.9107&0.9627&0.9525\\
 3&0.9185&0.9177&1.0006&0.9879\\
 4&0.9254&0.9245&1.0039&1.0060\\
 5&0.9308&0.9278&0.9977&0.9278\\
 \end{tabular}
\end{ruledtabular}
\caption{\label{tab:Areas_2}The (scaled) area (in units of lattice constant) under the steps in the local staircase structure obtained from the numerical results for the cumulative distribution function at times $t=5000,10000$ (in units of $1/g_1$) for $g=1/2$ $(> g_{c})$ near the external and internal extremal fronts (Figs.~\ref{fig:stair_cum_prob_e},~\ref{fig:stair_cum_prob_f}). The (scaled) area below each step near the external front and internal front remains nearly a unit constant. 
}
\end{table}
  
    When $g=g_{c}$, there is an additional internal extremal front at the origin. From our numerical calculations, we do not observe any staircase structure near the internal front at the origin. There is only a single sharp step of zero width with an anomalous $t^{1/4}$ scaling behaviour of $\delta \Phi$ for $g=g_c$ at the origin as shown in Fig. \ref{fig:stair_cum_prob_d}. Again, the numerical results are consistent with our theoretical analysis.  
For $g>g_c$, there are two additional first order extremal fronts occuring inside the causal region. The local scaling behaviour and staircase structure at these internal fronts, as shown in Fig.~\ref{fig:stair_cum_prob_f}, is again of the Airy type similar to that near the maximal fronts. However, since the extremal front velocities are different at the maximal and internal fronts, the height and width of the steps near the external and internal fronts are different as can be seen from  Fig. \ref{fig:stair_cum_prob_e} and Fig. \ref{fig:stair_cum_prob_f} for $g=1/2$. The (scaled) area under each step, however, remains a unit constant as shown in Table~\ref{tab:Areas_2} (right column).  

 Thus we find that the local scaling near the maximal fronts and in general near a first order extremal front is of the Airy type and leads to an internal staircase structure in the cumulative distribution profiles. There is no internal staircase structure near the second order front. As already pointed out, the global and local scaling properties of the cumulative distribution function for the simple QW (with only NN hopping) correspond exactly to that obtained for  magnetization profiles of  time-evolved domain wall states in the Ising chain \citep{antal_pre59, sasvari_pre69}. The Ising chain Hamiltonian can be transformed to a NN TB Hamiltonian by a Jordan-Wigner transformation \citep{Jordan-wigner}. Since the simple QW with only NN hopping is an alternative view of the NN TB Hamiltonian, this indicates that the time evolution of a single particle simple QW on the one dimensional lattice with an initially localized state is equivalent to the time evolution of a domain wall initial state in the Ising spin chain system. The integrable XX spin chain Hamiltonian  with three spin interactions of the XZX + YZY type \citep{suzuki,*suzuki2} given by:
 \begin{eqnarray}\label{eq:3spin_ham}
   H&&=\sum_i [J(S^{x}_i S^{x}_{i+1}+S^{y}_i S^{y}_{i+1}) \nonumber \\
   && \qquad \qquad +K(S^{x}_i S^{z}_{i+1}S^{x}_{i+2}+S^{y}_i S^{z}_{i+1} S^{y}_{i+2})]
  \end{eqnarray}
  can be transformed using a Jordan-Wigner transformation to the tight binding  Hamiltonian for spinless fermions with nearest and next-nearest neighbour hopping~\citep{titvinidze, pradeep2} and hence can be related to the QW model with NN and NNN hopping. We suggest that the cumulative distribution profiles for such a QW model with an initial localized state should correspond to the magnetization profiles of an initial domain wall state time-evolved according to the Hamiltonian Eq.~\ref{eq:3spin_ham}. The above spin chain model is known to have a phase diagram with two phases: one corresponding to a phase with one critical spin liquid and the other to a phase with two critical spin liquids with the transition occurring at a certain critical three spin interaction strength~\citep{titvinidze}. The observed two phases in the QW model with NN and NNN hopping can be connected to the two different phases in the above integrable spin chain model. We expect that the time evolved magnetization profiles of initial domain wall states in the spin system to also show a transition from a phase with one causal cone and internal staircase structure to a phase with two kinds of global and local scalings, i.e, show the existence of an internal light-cone and more than one kind of propagating quasiparticle. The magnetization profile at the critical $g$ value must show a second order extremal front at the origin with an anomalous $t ^{-1/2}$ scaling near the front. We note that the value of the critical NNN hopping  strength at which the transition occurs is different in the spin chain system and the quantum walk system (although they can be related). In the spin chain system, the transition occurs at the critical strength at which the number of Fermi points change from two to four while in the QW model, the transition occurs at the strength at which the number of extremal fronts change from two to four. 
   
 We also note here that the anomalous magnetization observed in the spin chain system~\citep{titvinidze} can be related to the difference in the total fraction of excitations with positive and negative $\lambda(q)$ and is given as:
\begin{equation}
m(g) = \left \{
  \begin{aligned}
    &\frac{1}{2}-\frac{q_e}{\pi} && \text{for}\ g\leq g_{c} \\
    &\frac{q_{i}- q_{e} }{\pi}-\frac{1}{2}  && \text{for}\  g\geq g_{c}\\
  \end{aligned} \right.
\end{equation}    
Thus the anomalous magnetization can be obtained from $q_e$ and $q_i$ or equivalently from the extremal front positions, $n_e$ and $n_i$. 

\section{\label{sec:localization}Measures of localization}
As already alluded to in the previous section, the lack of an internal staircase structure near the second order internal front at the critical NNN hopping strength points to localization. In this section, we study the localization properties of the time evolved wave packet by studying various measures characterizing the spreading and localization of the wave packet. In particular, we discuss   the inverse participation ratio, Shannon entropy and the return probability~\citep{cuevas,ipr_appl,Bera,QW_Krapivsky1,Krapivsky4}. The inverse participation ratio is used to characterize topological phases (like topological-band insulator transitions occurring in 2D Dirac materials like silicene), Anderson metal-insulator transition, many-body localization \citep{ipr_appl, Bera}, etc. Shannon entropy plays an important role in information theory and clarifies how much information is contained in the system \citep{cuevas}. The return probability or Loschmidt echo is also used to describe the localization of the wave packet~\citep{Krapivsky4}.

The inverse participation ratio ($\mbox{IPR}$) quantifies how many states a particle is distributed over when there is an uncertainty in the particle's position and hence is  the measure of localization of particle's wavefunction. It is defined as: 
 \begin{equation}\label{eq:ipr}
\mbox{IPR}(t)=‎‎\sum_{n=-\infty}^{\infty‎}p_{n}^{2}(t)
\end{equation}
 where $p_{n}(t)=|\psi(n,t)|^2$ is the probability of the particle to be at the site $n$ at a given instant of time $t$.
\begin{figure}[htp]           
\centering
\subfigure{\label{fig:ipr_a}}{\includegraphics[width=8cm,height=6cm,keepaspectratio]{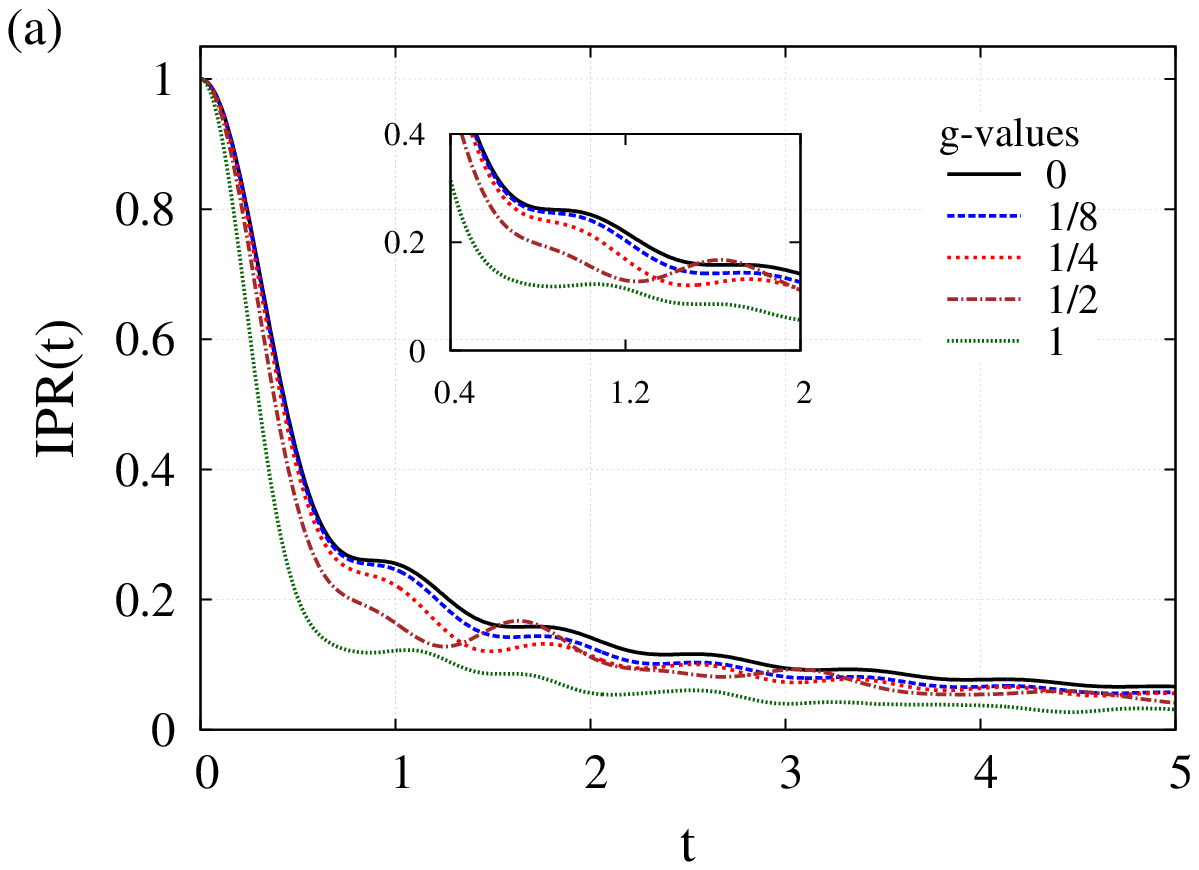}}\quad

\subfigure{\label{fig:ipr_g}}{\includegraphics[width=8cm,height=6cm,keepaspectratio]{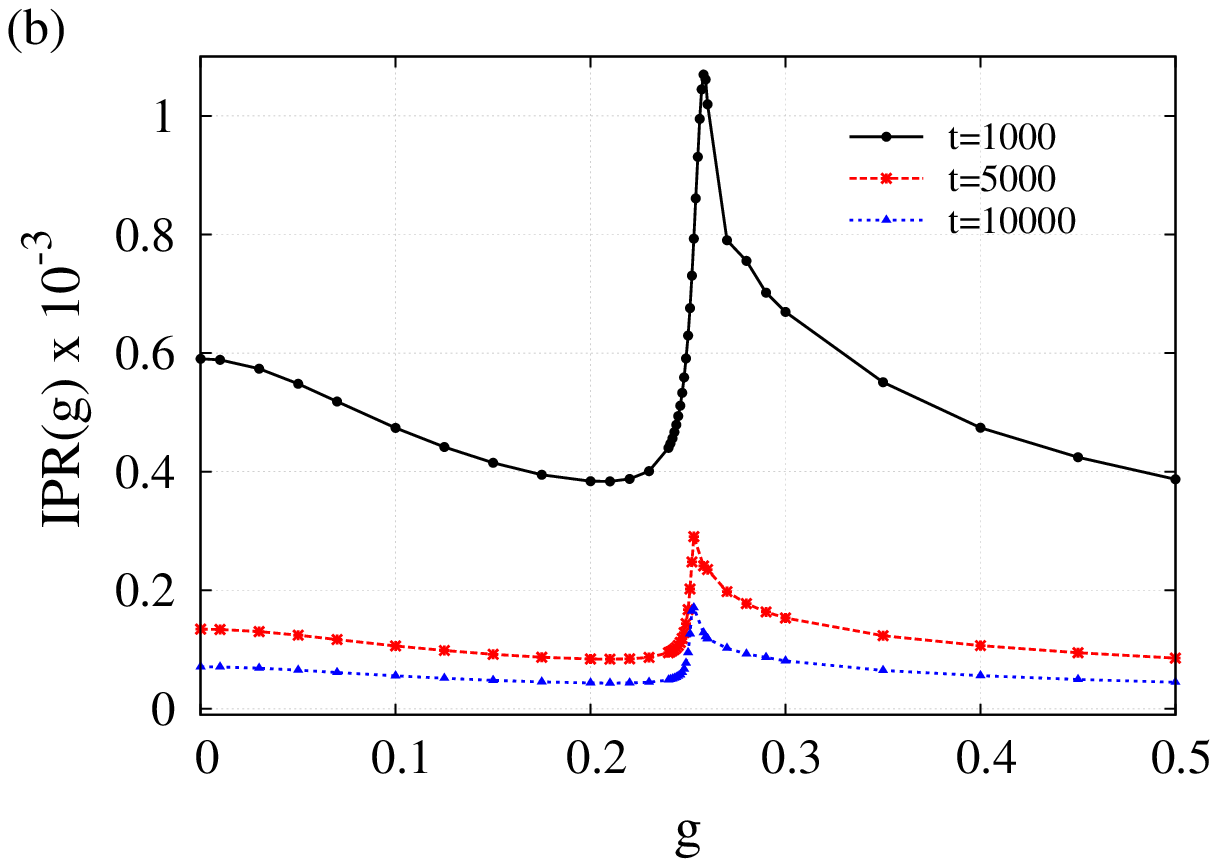}}
\caption{ The figure depicts the behaviour of the inverse participation ratio as obtained from  exact numerical calculations. Panel (a) shows the inverse participation ratio as a function of time for different values of NNN coupling $g$ for small $t$ values. The plot shows a decreasing amplitude with small oscillations (shown in the inset). Panel (b) shows the $g$ dependence of the inverse participation ratio for large times.  A peak is observed at $g=g_c=1/4$. (Time is measured in units of $1/g_1$.)}  \label{fig:ipr_gt}   
\end{figure}
 If a particle is localized at only one site, then $p_{n}=1$ for that site $n$, so $\mbox{IPR} = 1$ (large $\mbox{IPR}$) whereas if the particle is evenly distributed over all sites, i.e, if \,$p_{n} = 1/\sqrt{N}$ ($N$ is the total number of sites), then $\mbox{IPR} = 1/N$ (small $\mbox{IPR}$). Thus a large $\mbox{IPR}$ indicates localization since the contribution to the $\mbox{IPR}$ is mostly from the sites where the probability is large. For our system, the particle is initially localized at the origin, hence $\mbox{IPR}=1$ at $t=0$. With increasing time, the wave packet spreads and the particle delocalizes on the lattice. Therefore, $\mbox{IPR}$ decreases with increasing time as can be seen in  Fig.~\ref{fig:ipr_a} where we plot $\mbox{IPR}$ as a function of time for different $g$ values. In Fig.~\ref{fig:ipr_g}, we show the $g$-dependence of the $\mbox{IPR}$ at large times. It can be seen that the $\mbox{IPR}$ shows a peak at $g=g_c$.
\begin{figure}[htp]           
  \centering
  \subfigure{\label{fig:entropy_a}}{\includegraphics[width=8cm,height=6cm,keepaspectratio]{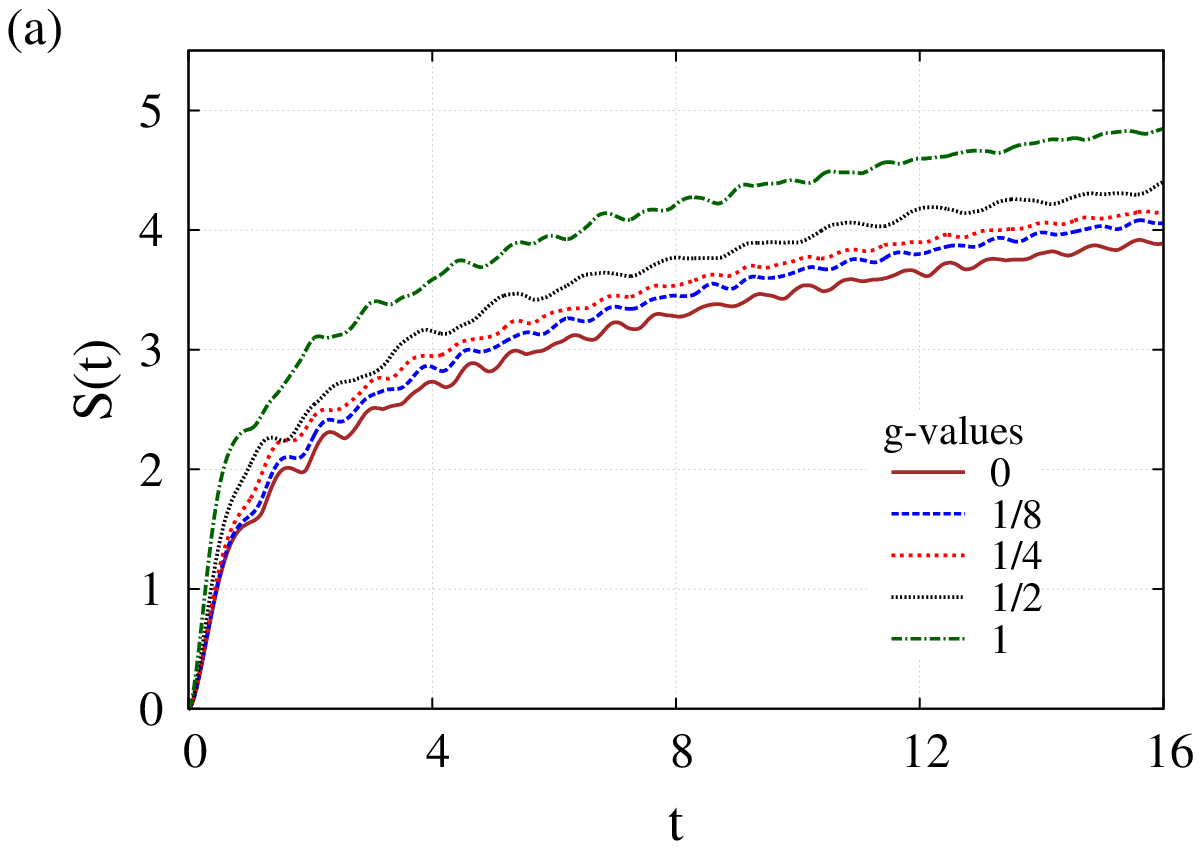}}\quad
  
  \subfigure{\label{fig:entropy_b}}{\includegraphics[width=8cm,height=6cm,keepaspectratio]{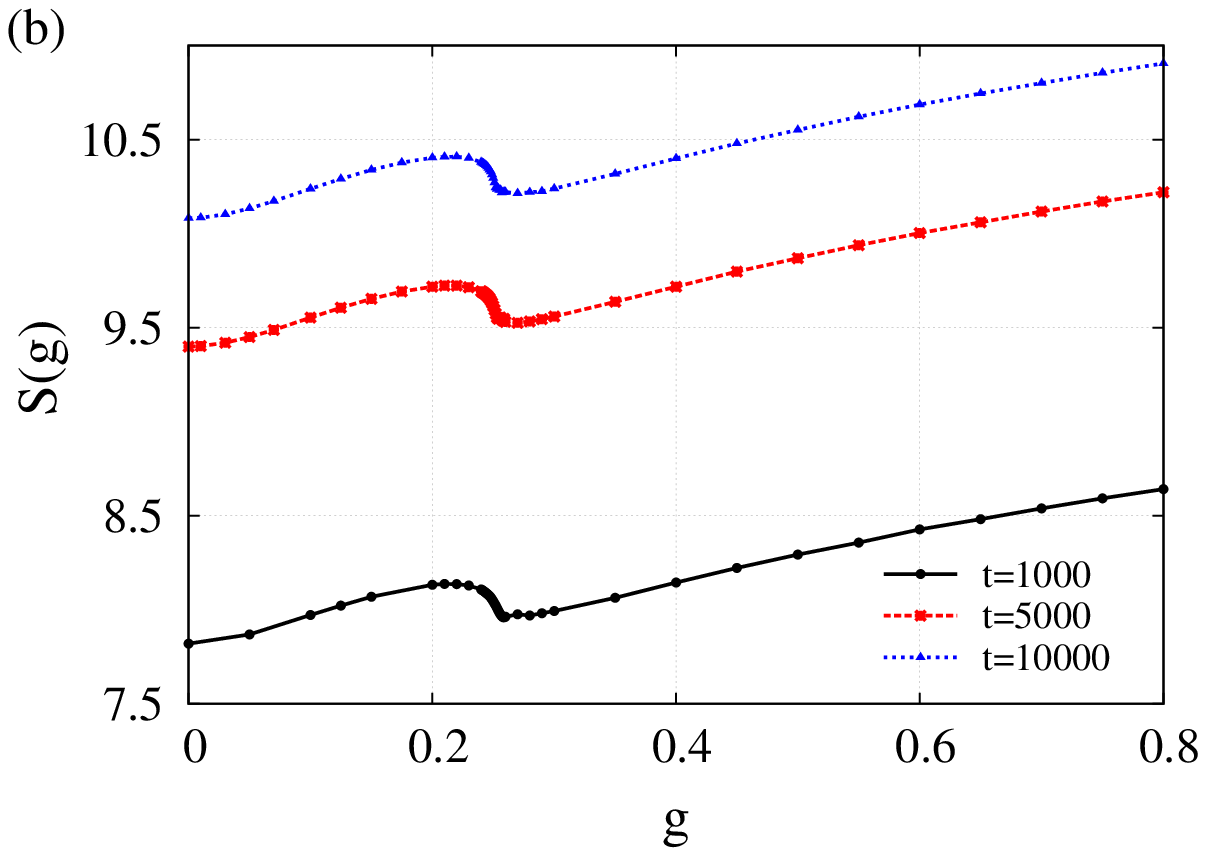}}
   \caption{Exact numerical results for the behaviour of the (dimensionless) Shannon entropy $S$ (a) as a function of time (measured in units of $1/g_1)$ for different values of NNN coupling strength $g$ for small times. The entropy can be seen to increase with time. (b) as a function of $g$ for large times. There is a dip in the entropy at $g =g_{c}$.  }  \label{fig:entropy}   
\end{figure}

  Another complementary measure of localization is the Shannon entropy $S$ defined as:
\begin{equation}\label{eq:entropy}
S(t)=-‎‎\sum_{n=-\infty}^{\infty‎}p_{n}(t)\,\ln \,{p_{n}(t)}
\end{equation}
The Shannon entropy  quantifies the amount of information present in the system. It also serves as a measure of localization because it is minimum (zero) when the particle is localized at one site. In Fig.~\ref{fig:entropy_a}, we plot the Shannon entropy as a function of time for different $g$ values. The $g$ dependence at large times is shown in Fig.~\ref{fig:entropy_b}. It can be seen from the Fig.~\ref{fig:entropy_a} that the entropy increases with time due to the increase in delocalization of the particle with time. We do not show this but the entropy grows logarithmically with time for asymptotically large times. From Fig.~\ref{fig:entropy_b}, we observe that the entropy shows a dip at $g=g_c$. 
Thus, the large time behaviour of the Shannon entropy and the inverse participation ratio both suggest that there is localization at $g=g_c$. 
\begin{figure}[htp]           
  \centering
  \subfigure{\label{fig:return_prob}}{\includegraphics[width=8cm,height=6cm,keepaspectratio]{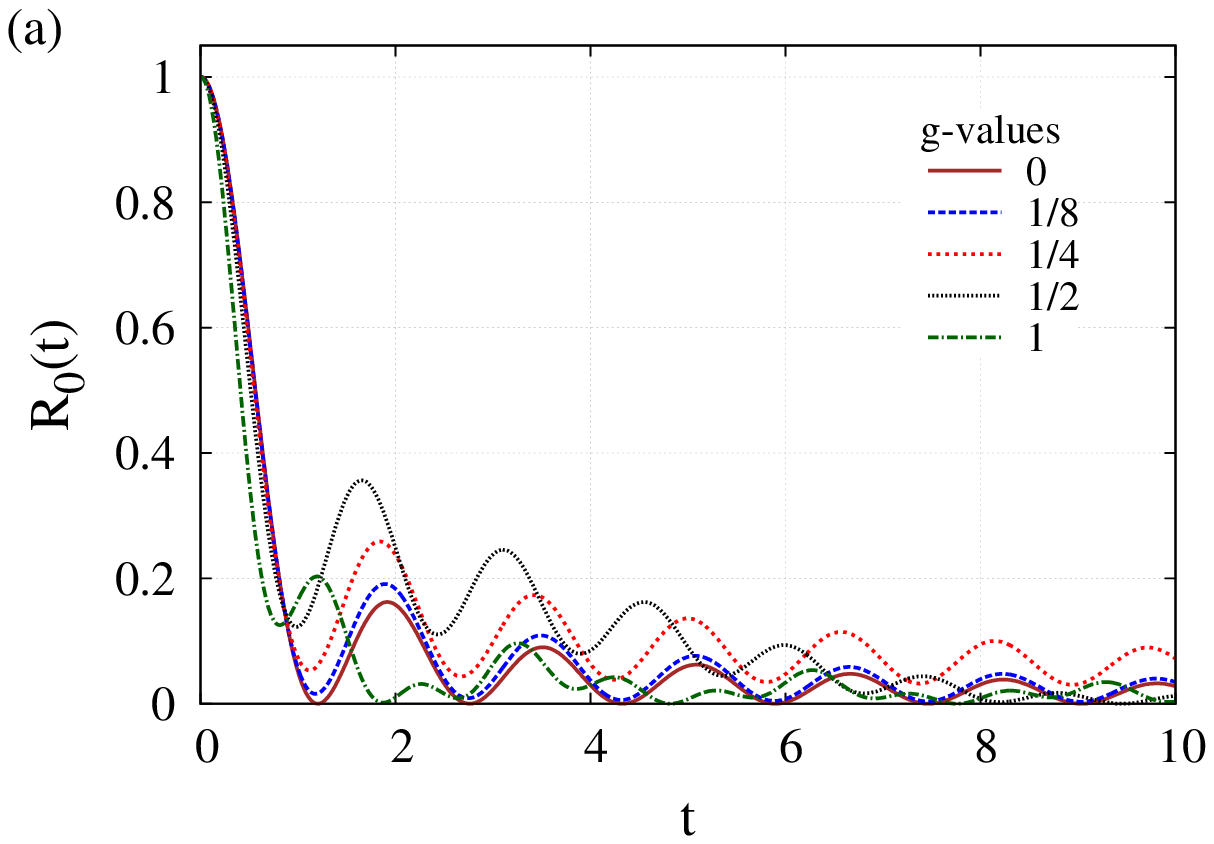}}\quad
  \subfigure{\label{fig:return_prob_g}}{\includegraphics[width=8cm,height=6cm,keepaspectratio]{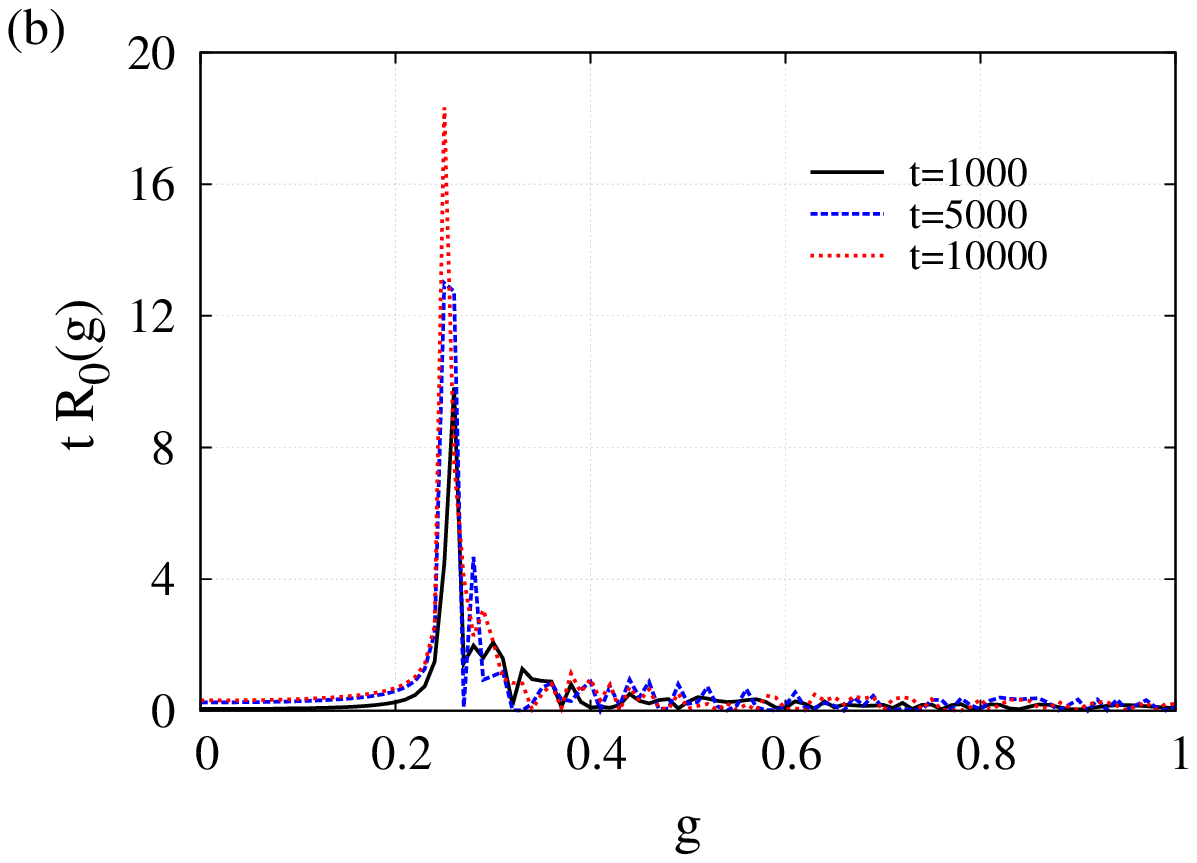}}
   \caption{ The plot showing the numerical results for the behaviour of the return probability $R_0$ (in units of the inverse lattice spacing $1/a$) at the origin as (a) a function of time (in units of $1/g_1$) for different values of NNN coupling strength $g$ at small times. $R_0$ shows in general an oscillatory behaviour with a decreasing amplitude. Aperiodic oscillations are observed for $g>g_c$. (b) The plot of $t R_0$ (in units of $1/{a g_1})$ showing the $g$ dependence of the return probability at large times. The plot shows non- monotonic $g$ dependence of the return probability with a peak at the critical NNN $g=g_{c}=1/4$ strength. At this value, there is always some finite probability for finding the particle at the origin.}  \label{fig:return}   
\end{figure} 

  We further confirm this by studying the return probability. The return probability is defined as the probability that the particle is found in the initial state  after some time $t $. For our case, where we assumed the particle to be initially localized at the origin, the return probability $R_0$ is given as:
 \begin{equation}\label{return_prob}
 R_0(t)=|\psi(0,t)|^2
 \end{equation}
  In the case of $g=0$, the return probability is simply given by $|J_{0}(v_e t)|^2$. Thus it has an oscillatory behaviour with an amplitude that decreases with time. In the presence of NNN hopping, one expects similar oscillatory behaviour for $g \neq g_c$, with an amplitude that decreases with increasing time. At the critical $g=g_c$, there is an additional internal front at the origin with a zero front velocity, i,e, there is a non-propagating excitation. Therefore we should expect that $R_0$ remains finite at all times leading to localization. The numerical results for the return probability are shown in Fig.~\ref{fig:return_prob} for different $g$ values for small times. The large time $g$ dependence of the return probability is shown in Fig.~\ref{fig:return_prob_g}. We can see from Fig.~\ref{fig:return_prob} that at small times, one finds in general, an oscillatory behaviour with an amplitude that decreases with  time. It can be seen Fig.~\ref{fig:return_prob} that for $g=g_c$, $R_0$ remains finite at all times. The aperiodic oscillations seen for $g>g_c$ can be attributed to the interference between the two kinds of propagating excitations. At large times, we can see from Fig.~\ref{fig:return_prob_g} that $R_0$ has a non-monotonic $g$ dependence with a peak at the critical value $g=g_{c}=1/4$. 
Or there is always some finite probability to find the particle at the origin. 
 
   Thus, the behaviour of all the different localization measures considered above, point to a localization transition at the critical $g$ value.

\section{\label{sec:conclusions}Conclusions}

We have studied the light-cone and front dynamics in a single particle extended QW model by studying the time evolution of an initially localized state. We showed that in general, propagating fronts can be either ordinary or extremal in nature. The latter exhibit anomalous sub-diffusive scaling behaviour in the front region. We investigated the dynamical scaling behaviour of the cumulative probability distribution function for the QW model with nearest and next-nearest neighbour hopping using analytic and numerical methods. The cumulative probability distribution function was shown to exhibit a dynamical 'causal light-cone'  inside which excitations propagate with a group velocity smaller than the maximal group velocity.  Excitations travelling faster than the maximal velocity are exponentially suppressed and lie outside the 'causal cone'.  Maximal fronts moving with a fixed 'light velocity' bound the causal cone. The front region spreads sub-diffusively with time with an Airy scaling that gives rise to a local staircase structure near the  maximal fronts. We interpret the existence of such a staircase structure as due to quantization of particle number.  At a certain critical NNN coupling strength, there is a transition from a phase with one 'causal cone' to a phase with two  nested 'causal cones' with the cumulative distribution profiles showing the emergence of an internal staircase structure inside the causal cone. The emergence of internal staircase structures suggests the existence of two different "velocities of light" and two kinds of quasiparticles. At the critical NNN coupling strength, a second order extremal front occurs at the  origin, exhibiting  $1/t^{1/4}$ sub-diffusive scaling behaviour in the front region, but does not show any local staircase structure. We also study the localization behaviour  of the wavefunction by studying various localization measures like the Shannon entropy, inverse participation ratio and the return probability. These studies point to a localization transition at the critical NNN hopping strength separating the two phases. The above study can be straightforwardly generalized to the case where the hopping is allowed upto a finite range of M neighbours. Depending on the relative strength of the various hopping parameters, one expects different phases characterized by m internal light-cones and where m can take values 0, 1, 2, . . . , M.

We also connect the present study to that of light-cone and front dynamics in spin chain systems by observing that the cumulative distribution profiles obtained for the simple QW (with only NN hopping) correspond exactly to the magnetization profiles of time-evolved domain wall states in the Ising chain \citep{antal_pre59, sasvari_pre69}.  We suggest that the time evolution of a single particle QW on the one dimensional lattice with an initially localized state  is equivalent to the time evolution of a domain wall initial state in the corresponding spin chain system. We note that recently, it was pointed out that the time evolution of an initial state with a single spin flip is the same as that of an initial domain wall state in the XXZ spin chain~\citep{Bulchandani}. We conjecture therefore that the cumulative distribution profiles for the QW model with nearest and next-nearest neighbour hopping with an initial localized state correspond to the magnetization profiles of an initial domain wall state time-evolving according to the integrable extended spin chain Hamiltonian  with three spin interactions of the XZX + YZY type. The study also provides an example of a model showing the emergence of internal ‘light-cones’ in nested integrable models with the existence of more than one kind of propagating quasiparticle. Finally, the study shows that the time dynamics of a single particle quantum walk on a lattice already captures the physics of a many body system and provides a different way of studying equilibrium and non-equilibrium many body physics theoretically and experimentally.

\acknowledgments{ P.D thanks DST India for financial support through research grant.} 

\bibliography{quantum_walk}

\begin{thebibliography}{43}%
\makeatletter
\providecommand \@ifxundefined [1]{%
 \@ifx{#1\undefined}
}%
\providecommand \@ifnum [1]{%
 \ifnum #1\expandafter \@firstoftwo
 \else \expandafter \@secondoftwo
 \fi
}%
\providecommand \@ifx [1]{%
 \ifx #1\expandafter \@firstoftwo
 \else \expandafter \@secondoftwo
 \fi
}%
\providecommand \natexlab [1]{#1}%
\providecommand \enquote  [1]{``#1''}%
\providecommand \bibnamefont  [1]{#1}%
\providecommand \bibfnamefont [1]{#1}%
\providecommand \citenamefont [1]{#1}%
\providecommand \href@noop [0]{\@secondoftwo}%
\providecommand \href [0]{\begingroup \@sanitize@url \@href}%
\providecommand \@href[1]{\@@startlink{#1}\@@href}%
\providecommand \@@href[1]{\endgroup#1\@@endlink}%
\providecommand \@sanitize@url [0]{\catcode `\\12\catcode `\$12\catcode
  `\&12\catcode `\#12\catcode `\^12\catcode `\_12\catcode `\%12\relax}%
\providecommand \@@startlink[1]{}%
\providecommand \@@endlink[0]{}%
\providecommand \url  [0]{\begingroup\@sanitize@url \@url }%
\providecommand \@url [1]{\endgroup\@href {#1}{\urlprefix }}%
\providecommand \urlprefix  [0]{URL }%
\providecommand \Eprint [0]{\href }%
\providecommand \doibase [0]{http://dx.doi.org/}%
\providecommand \selectlanguage [0]{\@gobble}%
\providecommand \bibinfo  [0]{\@secondoftwo}%
\providecommand \bibfield  [0]{\@secondoftwo}%
\providecommand \translation [1]{[#1]}%
\providecommand \BibitemOpen [0]{}%
\providecommand \bibitemStop [0]{}%
\providecommand \bibitemNoStop [0]{.\EOS\space}%
\providecommand \EOS [0]{\spacefactor3000\relax}%
\providecommand \BibitemShut  [1]{\csname bibitem#1\endcsname}%
\let\auto@bib@innerbib\@empty
\bibitem [{\citenamefont {Aharonov}\ \emph {et~al.}(1993)\citenamefont
  {Aharonov}, \citenamefont {Davidovich},\ and\ \citenamefont
  {Zagury}}]{Aharonov}%
  \BibitemOpen
  \bibfield  {author} {\bibinfo {author} {\bibfnamefont {Y.}~\bibnamefont
  {Aharonov}}, \bibinfo {author} {\bibfnamefont {L.}~\bibnamefont
  {Davidovich}}, \ and\ \bibinfo {author} {\bibfnamefont {N.}~\bibnamefont
  {Zagury}},\ }\href {\doibase 10.1103/PhysRevA.48.1687} {\bibfield  {journal}
  {\bibinfo  {journal} {Phys. Rev. A}\ }\textbf {\bibinfo {volume} {48}},\
  \bibinfo {pages} {1687} (\bibinfo {year} {1993})}\BibitemShut {NoStop}%
\bibitem [{\citenamefont {Kempe}(2003)}]{Kempe}%
  \BibitemOpen
  \bibfield  {author} {\bibinfo {author} {\bibfnamefont {J.}~\bibnamefont
  {Kempe}},\ }\href {\doibase 10.1080/00107151031000110776} {\bibfield
  {journal} {\bibinfo  {journal} {Contemporary Physics}\ }\textbf {\bibinfo
  {volume} {44}},\ \bibinfo {pages} {307} (\bibinfo {year} {2003})},\ \Eprint
  {http://arxiv.org/abs/https://doi.org/10.1080/00107151031000110776}
  {https://doi.org/10.1080/00107151031000110776} \BibitemShut {NoStop}%
\bibitem [{\citenamefont {Venegas-Andraca}(2012)}]{VAndraca}%
  \BibitemOpen
  \bibfield  {author} {\bibinfo {author} {\bibfnamefont {S.~E.}\ \bibnamefont
  {Venegas-Andraca}},\ }\href {\doibase 10.1007/s11128-012-0432-5} {\bibfield
  {journal} {\bibinfo  {journal} {Quantum Information Processing}\ }\textbf
  {\bibinfo {volume} {11}},\ \bibinfo {pages} {1015} (\bibinfo {year}
  {2012})}\BibitemShut {NoStop}%
\bibitem [{\citenamefont {M\"ulken}\ and\ \citenamefont
  {Blumen}(2011)}]{mulken}%
  \BibitemOpen
  \bibfield  {author} {\bibinfo {author} {\bibfnamefont {O.}~\bibnamefont
  {M\"ulken}}\ and\ \bibinfo {author} {\bibfnamefont {A.}~\bibnamefont
  {Blumen}},\ }\href {\doibase https://doi.org/10.1016/j.physrep.2011.01.002}
  {\bibfield  {journal} {\bibinfo  {journal} {Physics Reports}\ }\textbf
  {\bibinfo {volume} {502}},\ \bibinfo {pages} {37 } (\bibinfo {year}
  {2011})}\BibitemShut {NoStop}%
\bibitem [{\citenamefont {Longhi}(2009)}]{longhi}%
  \BibitemOpen
  \bibfield  {author} {\bibinfo {author} {\bibfnamefont {S.}~\bibnamefont
  {Longhi}},\ }\href {\doibase 10.1002/lpor.200810055} {\bibfield  {journal}
  {\bibinfo  {journal} {Laser \& Photonics Reviews}\ }\textbf {\bibinfo
  {volume} {3}},\ \bibinfo {pages} {243} (\bibinfo {year} {2009})}\BibitemShut
  {NoStop}%
\bibitem [{\citenamefont {Dreisow}\ \emph {et~al.}(2010)\citenamefont
  {Dreisow}, \citenamefont {Heinrich}, \citenamefont {Keil}, \citenamefont
  {T\"unnermann}, \citenamefont {Nolte}, \citenamefont {Longhi},\ and\
  \citenamefont {Szameit}}]{QW_Zitterberwegung}%
  \BibitemOpen
  \bibfield  {author} {\bibinfo {author} {\bibfnamefont {F.}~\bibnamefont
  {Dreisow}}, \bibinfo {author} {\bibfnamefont {M.}~\bibnamefont {Heinrich}},
  \bibinfo {author} {\bibfnamefont {R.}~\bibnamefont {Keil}}, \bibinfo {author}
  {\bibfnamefont {A.}~\bibnamefont {T\"unnermann}}, \bibinfo {author}
  {\bibfnamefont {S.}~\bibnamefont {Nolte}}, \bibinfo {author} {\bibfnamefont
  {S.}~\bibnamefont {Longhi}}, \ and\ \bibinfo {author} {\bibfnamefont
  {A.}~\bibnamefont {Szameit}},\ }\href {\doibase
  10.1103/PhysRevLett.105.143902} {\bibfield  {journal} {\bibinfo  {journal}
  {Phys. Rev. Lett.}\ }\textbf {\bibinfo {volume} {105}},\ \bibinfo {pages}
  {143902} (\bibinfo {year} {2010})}\BibitemShut {NoStop}%
\bibitem [{\citenamefont {Dreisow}\ \emph {et~al.}(2012)\citenamefont
  {Dreisow}, \citenamefont {Keil}, \citenamefont {Tünnermann}, \citenamefont
  {Nolte}, \citenamefont {Longhi},\ and\ \citenamefont
  {Szameit}}]{QW_Klein_tunneling}%
  \BibitemOpen
  \bibfield  {author} {\bibinfo {author} {\bibfnamefont {F.}~\bibnamefont
  {Dreisow}}, \bibinfo {author} {\bibfnamefont {R.}~\bibnamefont {Keil}},
  \bibinfo {author} {\bibfnamefont {A.}~\bibnamefont {Tünnermann}}, \bibinfo
  {author} {\bibfnamefont {S.}~\bibnamefont {Nolte}}, \bibinfo {author}
  {\bibfnamefont {S.}~\bibnamefont {Longhi}}, \ and\ \bibinfo {author}
  {\bibfnamefont {A.}~\bibnamefont {Szameit}},\ }\href
  {http://stacks.iop.org/0295-5075/97/i=1/a=10008} {\bibfield  {journal}
  {\bibinfo  {journal} {EPL}\ }\textbf {\bibinfo {volume} {97}},\ \bibinfo
  {pages} {10008} (\bibinfo {year} {2012})}\BibitemShut {NoStop}%
\bibitem [{\citenamefont {Naether}\ \emph {et~al.}(2012)\citenamefont
  {Naether}, \citenamefont {Meyer}, \citenamefont {St\"{u}tzer}, \citenamefont
  {T\"{u}nnermann}, \citenamefont {Nolte}, \citenamefont {Molina},\ and\
  \citenamefont {Szameit}}]{QW_Anderson_localization}%
  \BibitemOpen
  \bibfield  {author} {\bibinfo {author} {\bibfnamefont {U.}~\bibnamefont
  {Naether}}, \bibinfo {author} {\bibfnamefont {J.~M.}\ \bibnamefont {Meyer}},
  \bibinfo {author} {\bibfnamefont {S.}~\bibnamefont {St\"{u}tzer}}, \bibinfo
  {author} {\bibfnamefont {A.}~\bibnamefont {T\"{u}nnermann}}, \bibinfo
  {author} {\bibfnamefont {S.}~\bibnamefont {Nolte}}, \bibinfo {author}
  {\bibfnamefont {M.~I.}\ \bibnamefont {Molina}}, \ and\ \bibinfo {author}
  {\bibfnamefont {A.}~\bibnamefont {Szameit}},\ }\href {\doibase
  10.1364/OL.37.000485} {\bibfield  {journal} {\bibinfo  {journal} {Opt.
  Lett.}\ }\textbf {\bibinfo {volume} {37}},\ \bibinfo {pages} {485} (\bibinfo
  {year} {2012})}\BibitemShut {NoStop}%
\bibitem [{\citenamefont {Kitagawa}\ \emph {et~al.}(2012)\citenamefont
  {Kitagawa}, \citenamefont {Broome}, \citenamefont {Fedrizzi}, \citenamefont
  {Rudner}, \citenamefont {Berg}, \citenamefont {Kassal}, \citenamefont
  {Aspuru-Guzik}, \citenamefont {Demler},\ and\ \citenamefont
  {White}}]{QW_Topological_phases}%
  \BibitemOpen
  \bibfield  {author} {\bibinfo {author} {\bibfnamefont {T.}~\bibnamefont
  {Kitagawa}}, \bibinfo {author} {\bibfnamefont {M.}~\bibnamefont {Broome}},
  \bibinfo {author} {\bibfnamefont {A.}~\bibnamefont {Fedrizzi}}, \bibinfo
  {author} {\bibfnamefont {M.}~\bibnamefont {Rudner}}, \bibinfo {author}
  {\bibfnamefont {E.}~\bibnamefont {Berg}}, \bibinfo {author} {\bibfnamefont
  {I.}~\bibnamefont {Kassal}}, \bibinfo {author} {\bibfnamefont
  {A.}~\bibnamefont {Aspuru-Guzik}}, \bibinfo {author} {\bibfnamefont
  {E.}~\bibnamefont {Demler}}, \ and\ \bibinfo {author} {\bibfnamefont
  {A.}~\bibnamefont {White}},\ }\href {\doibase 10.1038/ncomms1872} {\bibfield
  {journal} {\bibinfo  {journal} {Nature Communications}\ }\textbf {\bibinfo
  {volume} {3}},\ \bibinfo {pages} {882 EP } (\bibinfo {year}
  {2012})}\BibitemShut {NoStop}%
\bibitem [{\citenamefont {Benedetti}\ \emph {et~al.}(2012)\citenamefont
  {Benedetti}, \citenamefont {Buscemi},\ and\ \citenamefont
  {Bordone}}]{QW_strong}%
  \BibitemOpen
  \bibfield  {author} {\bibinfo {author} {\bibfnamefont {C.}~\bibnamefont
  {Benedetti}}, \bibinfo {author} {\bibfnamefont {F.}~\bibnamefont {Buscemi}},
  \ and\ \bibinfo {author} {\bibfnamefont {P.}~\bibnamefont {Bordone}},\ }\href
  {\doibase 10.1103/PhysRevA.85.042314} {\bibfield  {journal} {\bibinfo
  {journal} {Phys. Rev. A}\ }\textbf {\bibinfo {volume} {85}},\ \bibinfo
  {pages} {042314} (\bibinfo {year} {2012})}\BibitemShut {NoStop}%
\bibitem [{\citenamefont {Qin}\ \emph {et~al.}(2014)\citenamefont {Qin},
  \citenamefont {Ke}, \citenamefont {Guan}, \citenamefont {Li}, \citenamefont
  {Andrei},\ and\ \citenamefont {Lee}}]{Nata_Andrei}%
  \BibitemOpen
  \bibfield  {author} {\bibinfo {author} {\bibfnamefont {X.}~\bibnamefont
  {Qin}}, \bibinfo {author} {\bibfnamefont {Y.}~\bibnamefont {Ke}}, \bibinfo
  {author} {\bibfnamefont {X.}~\bibnamefont {Guan}}, \bibinfo {author}
  {\bibfnamefont {Z.}~\bibnamefont {Li}}, \bibinfo {author} {\bibfnamefont
  {N.}~\bibnamefont {Andrei}}, \ and\ \bibinfo {author} {\bibfnamefont
  {C.}~\bibnamefont {Lee}},\ }\href {\doibase 10.1103/PhysRevA.90.062301}
  {\bibfield  {journal} {\bibinfo  {journal} {Phys. Rev. A}\ }\textbf {\bibinfo
  {volume} {90}},\ \bibinfo {pages} {062301} (\bibinfo {year}
  {2014})}\BibitemShut {NoStop}%
\bibitem [{\citenamefont {Z\"ahringer}\ \emph {et~al.}(2010)\citenamefont
  {Z\"ahringer}, \citenamefont {Kirchmair}, \citenamefont {Gerritsma},
  \citenamefont {Solano}, \citenamefont {Blatt},\ and\ \citenamefont
  {Roos}}]{QW_trapped_ion}%
  \BibitemOpen
  \bibfield  {author} {\bibinfo {author} {\bibfnamefont {F.}~\bibnamefont
  {Z\"ahringer}}, \bibinfo {author} {\bibfnamefont {G.}~\bibnamefont
  {Kirchmair}}, \bibinfo {author} {\bibfnamefont {R.}~\bibnamefont
  {Gerritsma}}, \bibinfo {author} {\bibfnamefont {E.}~\bibnamefont {Solano}},
  \bibinfo {author} {\bibfnamefont {R.}~\bibnamefont {Blatt}}, \ and\ \bibinfo
  {author} {\bibfnamefont {C.~F.}\ \bibnamefont {Roos}},\ }\href {\doibase
  10.1103/PhysRevLett.104.100503} {\bibfield  {journal} {\bibinfo  {journal}
  {Phys. Rev. Lett.}\ }\textbf {\bibinfo {volume} {104}},\ \bibinfo {pages}
  {100503} (\bibinfo {year} {2010})}\BibitemShut {NoStop}%
\bibitem [{\citenamefont {Travaglione}\ and\ \citenamefont
  {Milburn}(2002)}]{PRA.65.032310}%
  \BibitemOpen
  \bibfield  {author} {\bibinfo {author} {\bibfnamefont {B.~C.}\ \bibnamefont
  {Travaglione}}\ and\ \bibinfo {author} {\bibfnamefont {G.~J.}\ \bibnamefont
  {Milburn}},\ }\href {\doibase 10.1103/PhysRevA.65.032310} {\bibfield
  {journal} {\bibinfo  {journal} {Phys. Rev. A}\ }\textbf {\bibinfo {volume}
  {65}},\ \bibinfo {pages} {032310} (\bibinfo {year} {2002})}\BibitemShut
  {NoStop}%
\bibitem [{\citenamefont {Schmitz}\ \emph {et~al.}(2009)\citenamefont
  {Schmitz}, \citenamefont {Matjeschk}, \citenamefont {Schneider},
  \citenamefont {Glueckert}, \citenamefont {Enderlein}, \citenamefont {Huber},\
  and\ \citenamefont {Schaetz}}]{PRL.103.090504}%
  \BibitemOpen
  \bibfield  {author} {\bibinfo {author} {\bibfnamefont {H.}~\bibnamefont
  {Schmitz}}, \bibinfo {author} {\bibfnamefont {R.}~\bibnamefont {Matjeschk}},
  \bibinfo {author} {\bibfnamefont {C.}~\bibnamefont {Schneider}}, \bibinfo
  {author} {\bibfnamefont {J.}~\bibnamefont {Glueckert}}, \bibinfo {author}
  {\bibfnamefont {M.}~\bibnamefont {Enderlein}}, \bibinfo {author}
  {\bibfnamefont {T.}~\bibnamefont {Huber}}, \ and\ \bibinfo {author}
  {\bibfnamefont {T.}~\bibnamefont {Schaetz}},\ }\href {\doibase
  10.1103/PhysRevLett.103.090504} {\bibfield  {journal} {\bibinfo  {journal}
  {Phys. Rev. Lett.}\ }\textbf {\bibinfo {volume} {103}},\ \bibinfo {pages}
  {090504} (\bibinfo {year} {2009})}\BibitemShut {NoStop}%
\bibitem [{\citenamefont {Sanders}\ \emph {et~al.}(2003)\citenamefont
  {Sanders}, \citenamefont {Bartlett}, \citenamefont {Tregenna},\ and\
  \citenamefont {Knight}}]{QW_QED_PhysRevA.67.042305}%
  \BibitemOpen
  \bibfield  {author} {\bibinfo {author} {\bibfnamefont {B.~C.}\ \bibnamefont
  {Sanders}}, \bibinfo {author} {\bibfnamefont {S.~D.}\ \bibnamefont
  {Bartlett}}, \bibinfo {author} {\bibfnamefont {B.}~\bibnamefont {Tregenna}},
  \ and\ \bibinfo {author} {\bibfnamefont {P.~L.}\ \bibnamefont {Knight}},\
  }\href {\doibase 10.1103/PhysRevA.67.042305} {\bibfield  {journal} {\bibinfo
  {journal} {Phys. Rev. A}\ }\textbf {\bibinfo {volume} {67}},\ \bibinfo
  {pages} {042305} (\bibinfo {year} {2003})}\BibitemShut {NoStop}%
\bibitem [{\citenamefont {Schreiber}\ \emph {et~al.}(2010)\citenamefont
  {Schreiber}, \citenamefont {Cassemiro}, \citenamefont
  {Poto\ifmmode~\check{c}\else \v{c}\fi{}ek}, \citenamefont {G\'abris},
  \citenamefont {Mosley}, \citenamefont {Andersson}, \citenamefont {Jex},\ and\
  \citenamefont {Silberhorn}}]{QW_photons}%
  \BibitemOpen
  \bibfield  {author} {\bibinfo {author} {\bibfnamefont {A.}~\bibnamefont
  {Schreiber}}, \bibinfo {author} {\bibfnamefont {K.~N.}\ \bibnamefont
  {Cassemiro}}, \bibinfo {author} {\bibfnamefont {V.}~\bibnamefont
  {Poto\ifmmode~\check{c}\else \v{c}\fi{}ek}}, \bibinfo {author} {\bibfnamefont
  {A.}~\bibnamefont {G\'abris}}, \bibinfo {author} {\bibfnamefont {P.~J.}\
  \bibnamefont {Mosley}}, \bibinfo {author} {\bibfnamefont {E.}~\bibnamefont
  {Andersson}}, \bibinfo {author} {\bibfnamefont {I.}~\bibnamefont {Jex}}, \
  and\ \bibinfo {author} {\bibfnamefont {C.}~\bibnamefont {Silberhorn}},\
  }\href {\doibase 10.1103/PhysRevLett.104.050502} {\bibfield  {journal}
  {\bibinfo  {journal} {Phys. Rev. Lett.}\ }\textbf {\bibinfo {volume} {104}},\
  \bibinfo {pages} {050502} (\bibinfo {year} {2010})}\BibitemShut {NoStop}%
\bibitem [{\citenamefont {D\"ur}\ \emph {et~al.}(2002)\citenamefont {D\"ur},
  \citenamefont {Raussendorf}, \citenamefont {Kendon},\ and\ \citenamefont
  {Briegel}}]{QW_optical_lattice}%
  \BibitemOpen
  \bibfield  {author} {\bibinfo {author} {\bibfnamefont {W.}~\bibnamefont
  {D\"ur}}, \bibinfo {author} {\bibfnamefont {R.}~\bibnamefont {Raussendorf}},
  \bibinfo {author} {\bibfnamefont {V.~M.}\ \bibnamefont {Kendon}}, \ and\
  \bibinfo {author} {\bibfnamefont {H.-J.}\ \bibnamefont {Briegel}},\ }\href
  {\doibase 10.1103/PhysRevA.66.052319} {\bibfield  {journal} {\bibinfo
  {journal} {Phys. Rev. A}\ }\textbf {\bibinfo {volume} {66}},\ \bibinfo
  {pages} {052319} (\bibinfo {year} {2002})}\BibitemShut {NoStop}%
\bibitem [{\citenamefont {Karski}\ \emph {et~al.}(2009)\citenamefont {Karski},
  \citenamefont {Förster}, \citenamefont {Choi}, \citenamefont {Steffen},
  \citenamefont {Alt}, \citenamefont {Meschede},\ and\ \citenamefont
  {Widera}}]{QW_trapped_atom}%
  \BibitemOpen
  \bibfield  {author} {\bibinfo {author} {\bibfnamefont {M.}~\bibnamefont
  {Karski}}, \bibinfo {author} {\bibfnamefont {L.}~\bibnamefont {Förster}},
  \bibinfo {author} {\bibfnamefont {J.-M.}\ \bibnamefont {Choi}}, \bibinfo
  {author} {\bibfnamefont {A.}~\bibnamefont {Steffen}}, \bibinfo {author}
  {\bibfnamefont {W.}~\bibnamefont {Alt}}, \bibinfo {author} {\bibfnamefont
  {D.}~\bibnamefont {Meschede}}, \ and\ \bibinfo {author} {\bibfnamefont
  {A.}~\bibnamefont {Widera}},\ }\href {\doibase 10.1126/science.1174436}
  {\bibfield  {journal} {\bibinfo  {journal} {Science (New York, N.Y.)}\
  }\textbf {\bibinfo {volume} {325}},\ \bibinfo {pages} {174} (\bibinfo {year}
  {2009})}\BibitemShut {NoStop}%
\bibitem [{\citenamefont {Dadras}\ \emph {et~al.}(2018)\citenamefont {Dadras},
  \citenamefont {Gresch}, \citenamefont {Groiseau}, \citenamefont {Wimberger},\
  and\ \citenamefont {Summy}}]{QW_BEC}%
  \BibitemOpen
  \bibfield  {author} {\bibinfo {author} {\bibfnamefont {S.}~\bibnamefont
  {Dadras}}, \bibinfo {author} {\bibfnamefont {A.}~\bibnamefont {Gresch}},
  \bibinfo {author} {\bibfnamefont {C.}~\bibnamefont {Groiseau}}, \bibinfo
  {author} {\bibfnamefont {S.}~\bibnamefont {Wimberger}}, \ and\ \bibinfo
  {author} {\bibfnamefont {G.~S.}\ \bibnamefont {Summy}},\ }\href {\doibase
  10.1103/PhysRevLett.121.070402} {\bibfield  {journal} {\bibinfo  {journal}
  {Phys. Rev. Lett.}\ }\textbf {\bibinfo {volume} {121}},\ \bibinfo {pages}
  {070402} (\bibinfo {year} {2018})}\BibitemShut {NoStop}%
\bibitem [{\citenamefont {Arias}\ and\ \citenamefont {Luck}(1998)}]{Toro}%
  \BibitemOpen
  \bibfield  {author} {\bibinfo {author} {\bibfnamefont {S.~D.~T.}\
  \bibnamefont {Arias}}\ and\ \bibinfo {author} {\bibfnamefont {J.~M.}\
  \bibnamefont {Luck}},\ }\href {http://stacks.iop.org/0305-4470/31/i=38/a=007}
  {\bibfield  {journal} {\bibinfo  {journal} {Journal of Physics A:
  Mathematical and General}\ }\textbf {\bibinfo {volume} {31}},\ \bibinfo
  {pages} {7699} (\bibinfo {year} {1998})}\BibitemShut {NoStop}%
\bibitem [{\citenamefont {Krapivsky}\ \emph {et~al.}(2014)\citenamefont
  {Krapivsky}, \citenamefont {Luck},\ and\ \citenamefont
  {Mallick}}]{QW_Krapivsky1}%
  \BibitemOpen
  \bibfield  {author} {\bibinfo {author} {\bibfnamefont {P.~L.}\ \bibnamefont
  {Krapivsky}}, \bibinfo {author} {\bibfnamefont {J.~M.}\ \bibnamefont {Luck}},
  \ and\ \bibinfo {author} {\bibfnamefont {K.}~\bibnamefont {Mallick}},\ }\href
  {\doibase 10.1007/s10955-014-0936-8} {\bibfield  {journal} {\bibinfo
  {journal} {Journal of Statistical Physics}\ }\textbf {\bibinfo {volume}
  {154}},\ \bibinfo {pages} {1430} (\bibinfo {year} {2014})}\BibitemShut
  {NoStop}%
\bibitem [{\citenamefont {Krapivsky}\ \emph {et~al.}(2015)\citenamefont
  {Krapivsky}, \citenamefont {Luck},\ and\ \citenamefont
  {Mallick}}]{QW_Krapivsky2}%
  \BibitemOpen
  \bibfield  {author} {\bibinfo {author} {\bibfnamefont {P.~L.}\ \bibnamefont
  {Krapivsky}}, \bibinfo {author} {\bibfnamefont {J.~M.}\ \bibnamefont {Luck}},
  \ and\ \bibinfo {author} {\bibfnamefont {K.}~\bibnamefont {Mallick}},\ }\href
  {http://stacks.iop.org/1751-8121/48/i=47/a=475301} {\bibfield  {journal}
  {\bibinfo  {journal} {Journal of Physics A: Mathematical and Theoretical}\
  }\textbf {\bibinfo {volume} {48}},\ \bibinfo {pages} {475301} (\bibinfo
  {year} {2015})}\BibitemShut {NoStop}%
\bibitem [{\citenamefont {Lieb}\ and\ \citenamefont
  {Robinson}(1972)}]{lieb-robinson}%
  \BibitemOpen
  \bibfield  {author} {\bibinfo {author} {\bibfnamefont {E.~H.}\ \bibnamefont
  {Lieb}}\ and\ \bibinfo {author} {\bibfnamefont {D.~W.}\ \bibnamefont
  {Robinson}},\ }\href {\doibase 10.1007/BF01645779} {\bibfield  {journal}
  {\bibinfo  {journal} {Communications in Mathematical Physics}\ }\textbf
  {\bibinfo {volume} {28}},\ \bibinfo {pages} {251} (\bibinfo {year}
  {1972})}\BibitemShut {NoStop}%
\bibitem [{\citenamefont {Bravyi}\ \emph {et~al.}(2006)\citenamefont {Bravyi},
  \citenamefont {Hastings},\ and\ \citenamefont {Verstraete}}]{bravyi2006}%
  \BibitemOpen
  \bibfield  {author} {\bibinfo {author} {\bibfnamefont {S.}~\bibnamefont
  {Bravyi}}, \bibinfo {author} {\bibfnamefont {M.~B.}\ \bibnamefont
  {Hastings}}, \ and\ \bibinfo {author} {\bibfnamefont {F.}~\bibnamefont
  {Verstraete}},\ }\href {\doibase 10.1103/PhysRevLett.97.050401} {\bibfield
  {journal} {\bibinfo  {journal} {Phys. Rev. Lett.}\ }\textbf {\bibinfo
  {volume} {97}},\ \bibinfo {pages} {050401} (\bibinfo {year}
  {2006})}\BibitemShut {NoStop}%
\bibitem [{\citenamefont {Calabrese}\ and\ \citenamefont
  {Cardy}(2006)}]{cardy2006}%
  \BibitemOpen
  \bibfield  {author} {\bibinfo {author} {\bibfnamefont {P.}~\bibnamefont
  {Calabrese}}\ and\ \bibinfo {author} {\bibfnamefont {J.}~\bibnamefont
  {Cardy}},\ }\href {\doibase 10.1103/PhysRevLett.96.136801} {\bibfield
  {journal} {\bibinfo  {journal} {Phys. Rev. Lett.}\ }\textbf {\bibinfo
  {volume} {96}},\ \bibinfo {pages} {136801} (\bibinfo {year}
  {2006})}\BibitemShut {NoStop}%
\bibitem [{\citenamefont {Antal}\ \emph {et~al.}(1999)\citenamefont {Antal},
  \citenamefont {R\'acz}, \citenamefont {R\'akos},\ and\ \citenamefont
  {Sch\"utz}}]{antal_pre59}%
  \BibitemOpen
  \bibfield  {author} {\bibinfo {author} {\bibfnamefont {T.}~\bibnamefont
  {Antal}}, \bibinfo {author} {\bibfnamefont {Z.}~\bibnamefont {R\'acz}},
  \bibinfo {author} {\bibfnamefont {A.}~\bibnamefont {R\'akos}}, \ and\
  \bibinfo {author} {\bibfnamefont {G.~M.}\ \bibnamefont {Sch\"utz}},\ }\href
  {\doibase 10.1103/PhysRevE.59.4912} {\bibfield  {journal} {\bibinfo
  {journal} {Phys. Rev. E}\ }\textbf {\bibinfo {volume} {59}},\ \bibinfo
  {pages} {4912} (\bibinfo {year} {1999})}\BibitemShut {NoStop}%
\bibitem [{\citenamefont {Hunyadi}\ \emph {et~al.}(2004)\citenamefont
  {Hunyadi}, \citenamefont {R\'acz},\ and\ \citenamefont
  {Sasv\'ari}}]{sasvari_pre69}%
  \BibitemOpen
  \bibfield  {author} {\bibinfo {author} {\bibfnamefont {V.}~\bibnamefont
  {Hunyadi}}, \bibinfo {author} {\bibfnamefont {Z.}~\bibnamefont {R\'acz}}, \
  and\ \bibinfo {author} {\bibfnamefont {L.}~\bibnamefont {Sasv\'ari}},\ }\href
  {\doibase 10.1103/PhysRevE.69.066103} {\bibfield  {journal} {\bibinfo
  {journal} {Phys. Rev. E}\ }\textbf {\bibinfo {volume} {69}},\ \bibinfo
  {pages} {066103} (\bibinfo {year} {2004})}\BibitemShut {NoStop}%
\bibitem [{\citenamefont {Bonnes}\ \emph {et~al.}(2014)\citenamefont {Bonnes},
  \citenamefont {Essler},\ and\ \citenamefont {L\"auchli}}]{bonnes2014}%
  \BibitemOpen
  \bibfield  {author} {\bibinfo {author} {\bibfnamefont {L.}~\bibnamefont
  {Bonnes}}, \bibinfo {author} {\bibfnamefont {F.~H.~L.}\ \bibnamefont
  {Essler}}, \ and\ \bibinfo {author} {\bibfnamefont {A.~M.}\ \bibnamefont
  {L\"auchli}},\ }\href {\doibase 10.1103/PhysRevLett.113.187203} {\bibfield
  {journal} {\bibinfo  {journal} {Phys. Rev. Lett.}\ }\textbf {\bibinfo
  {volume} {113}},\ \bibinfo {pages} {187203} (\bibinfo {year}
  {2014})}\BibitemShut {NoStop}%
\bibitem [{\citenamefont {Cheneau}\ \emph {et~al.}(2012)\citenamefont
  {Cheneau}, \citenamefont {Barmettler}, \citenamefont {Poletti}, \citenamefont
  {Endres}, \citenamefont {Schau{\ss}}, \citenamefont {Fukuhara}, \citenamefont
  {Gross}, \citenamefont {Bloch}, \citenamefont {Kollath},\ and\ \citenamefont
  {Kuhr}}]{cheneau2012}%
  \BibitemOpen
  \bibfield  {author} {\bibinfo {author} {\bibfnamefont {M.}~\bibnamefont
  {Cheneau}}, \bibinfo {author} {\bibfnamefont {P.}~\bibnamefont {Barmettler}},
  \bibinfo {author} {\bibfnamefont {D.}~\bibnamefont {Poletti}}, \bibinfo
  {author} {\bibfnamefont {M.}~\bibnamefont {Endres}}, \bibinfo {author}
  {\bibfnamefont {P.}~\bibnamefont {Schau{\ss}}}, \bibinfo {author}
  {\bibfnamefont {T.}~\bibnamefont {Fukuhara}}, \bibinfo {author}
  {\bibfnamefont {C.}~\bibnamefont {Gross}}, \bibinfo {author} {\bibfnamefont
  {I.}~\bibnamefont {Bloch}}, \bibinfo {author} {\bibfnamefont
  {C.}~\bibnamefont {Kollath}}, \ and\ \bibinfo {author} {\bibfnamefont
  {S.}~\bibnamefont {Kuhr}},\ }\href {https://doi.org/10.1038/nature10748}
  {\bibfield  {journal} {\bibinfo  {journal} {Nature}\ }\textbf {\bibinfo
  {volume} {481}},\ \bibinfo {pages} {484 EP } (\bibinfo {year}
  {2012})}\BibitemShut {NoStop}%
\bibitem [{\citenamefont {Misguich}\ \emph {et~al.}(2017)\citenamefont
  {Misguich}, \citenamefont {Mallick},\ and\ \citenamefont
  {Krapivsky}}]{Krapivsky3}%
  \BibitemOpen
  \bibfield  {author} {\bibinfo {author} {\bibfnamefont {G.}~\bibnamefont
  {Misguich}}, \bibinfo {author} {\bibfnamefont {K.}~\bibnamefont {Mallick}}, \
  and\ \bibinfo {author} {\bibfnamefont {P.~L.}\ \bibnamefont {Krapivsky}},\
  }\href {\doibase 10.1103/PhysRevB.96.195151} {\bibfield  {journal} {\bibinfo
  {journal} {Phys. Rev. B}\ }\textbf {\bibinfo {volume} {96}},\ \bibinfo
  {pages} {195151} (\bibinfo {year} {2017})}\BibitemShut {NoStop}%
\bibitem [{\citenamefont {Collura}\ \emph {et~al.}(2018)\citenamefont
  {Collura}, \citenamefont {De~Luca},\ and\ \citenamefont {Viti}}]{collura}%
  \BibitemOpen
  \bibfield  {author} {\bibinfo {author} {\bibfnamefont {M.}~\bibnamefont
  {Collura}}, \bibinfo {author} {\bibfnamefont {A.}~\bibnamefont {De~Luca}}, \
  and\ \bibinfo {author} {\bibfnamefont {J.}~\bibnamefont {Viti}},\ }\href
  {\doibase 10.1103/PhysRevB.97.081111} {\bibfield  {journal} {\bibinfo
  {journal} {Phys. Rev. B}\ }\textbf {\bibinfo {volume} {97}},\ \bibinfo
  {pages} {081111} (\bibinfo {year} {2018})}\BibitemShut {NoStop}%
\bibitem [{\citenamefont {Najafi}\ \emph {et~al.}(2018)\citenamefont {Najafi},
  \citenamefont {Rajabpour},\ and\ \citenamefont {Viti}}]{najafi2018}%
  \BibitemOpen
  \bibfield  {author} {\bibinfo {author} {\bibfnamefont {K.}~\bibnamefont
  {Najafi}}, \bibinfo {author} {\bibfnamefont {M.~A.}\ \bibnamefont
  {Rajabpour}}, \ and\ \bibinfo {author} {\bibfnamefont {J.}~\bibnamefont
  {Viti}},\ }\href {\doibase 10.1103/PhysRevB.97.205103} {\bibfield  {journal}
  {\bibinfo  {journal} {Phys. Rev. B}\ }\textbf {\bibinfo {volume} {97}},\
  \bibinfo {pages} {205103} (\bibinfo {year} {2018})}\BibitemShut {NoStop}%
\bibitem [{\citenamefont {Jordan}\ and\ \citenamefont
  {Wigner}(1928)}]{Jordan-wigner}%
  \BibitemOpen
  \bibfield  {author} {\bibinfo {author} {\bibfnamefont {P.}~\bibnamefont
  {Jordan}}\ and\ \bibinfo {author} {\bibfnamefont {E.}~\bibnamefont
  {Wigner}},\ }\href {\doibase 10.1007/BF01331938} {\bibfield  {journal}
  {\bibinfo  {journal} {Zeitschrift f{\"u}r Physik}\ }\textbf {\bibinfo
  {volume} {47}},\ \bibinfo {pages} {631} (\bibinfo {year} {1928})}\BibitemShut
  {NoStop}%
\bibitem [{\citenamefont {Suzuki}(1971{\natexlab{a}})}]{suzuki}%
  \BibitemOpen
  \bibfield  {author} {\bibinfo {author} {\bibfnamefont {M.}~\bibnamefont
  {Suzuki}},\ }\href {\doibase https://doi.org/10.1016/0375-9601(71)90218-0}
  {\bibfield  {journal} {\bibinfo  {journal} {Physics Letters A}\ }\textbf
  {\bibinfo {volume} {34}},\ \bibinfo {pages} {94 } (\bibinfo {year}
  {1971}{\natexlab{a}})}\BibitemShut {NoStop}%
\bibitem [{\citenamefont {Suzuki}(1971{\natexlab{b}})}]{suzuki2}%
  \BibitemOpen
  \bibfield  {author} {\bibinfo {author} {\bibfnamefont {M.}~\bibnamefont
  {Suzuki}},\ }\href {\doibase 10.1143/PTP.46.1337} {\bibfield  {journal}
  {\bibinfo  {journal} {Progress of Theoretical Physics}\ }\textbf {\bibinfo
  {volume} {46}},\ \bibinfo {pages} {1337} (\bibinfo {year}
  {1971}{\natexlab{b}})}\BibitemShut {NoStop}%
\bibitem [{\citenamefont {Titvinidze}\ and\ \citenamefont
  {Japaridze}(2003)}]{titvinidze}%
  \BibitemOpen
  \bibfield  {author} {\bibinfo {author} {\bibfnamefont {I.}~\bibnamefont
  {Titvinidze}}\ and\ \bibinfo {author} {\bibfnamefont {G.~I.}\ \bibnamefont
  {Japaridze}},\ }\href {\doibase 10.1140/epjb/e2003-00113-8} {\bibfield
  {journal} {\bibinfo  {journal} {Eur. Phys. J. B}\ }\textbf {\bibinfo {volume}
  {32}},\ \bibinfo {pages} {383} (\bibinfo {year} {2003})}\BibitemShut
  {NoStop}%
\bibitem [{\citenamefont {{Cuevas}}\ \emph {et~al.}(2011)\citenamefont
  {{Cuevas}}, \citenamefont {{Curilef}},\ and\ \citenamefont
  {{Plastino}}}]{cuevas}%
  \BibitemOpen
  \bibfield  {author} {\bibinfo {author} {\bibfnamefont {F.~A.}\ \bibnamefont
  {{Cuevas}}}, \bibinfo {author} {\bibfnamefont {S.}~\bibnamefont {{Curilef}}},
  \ and\ \bibinfo {author} {\bibfnamefont {A.~R.}\ \bibnamefont {{Plastino}}},\
  }\href {\doibase 10.1016/j.aop.2011.07.003} {\bibfield  {journal} {\bibinfo
  {journal} {Annals of Physics}\ }\textbf {\bibinfo {volume} {326}},\ \bibinfo
  {pages} {2834} (\bibinfo {year} {2011})}\BibitemShut {NoStop}%
\bibitem [{\citenamefont {Abramowitz}\ and\ \citenamefont
  {Stegun}(1964)}]{abramowitz+stegun}%
  \BibitemOpen
  \bibfield  {author} {\bibinfo {author} {\bibfnamefont {M.}~\bibnamefont
  {Abramowitz}}\ and\ \bibinfo {author} {\bibfnamefont {I.~A.}\ \bibnamefont
  {Stegun}},\ }\href@noop {} {\emph {\bibinfo {title} {Handbook of Mathematical
  Functions with Formulas, Graphs, and Mathematical Tables}}},\ \bibinfo
  {edition} {ninth dover printing, tenth gpo printing}\ ed.\ (\bibinfo
  {publisher} {Dover},\ \bibinfo {address} {New York},\ \bibinfo {year}
  {1964})\BibitemShut {NoStop}%
\bibitem [{\citenamefont {Thakur}\ and\ \citenamefont
  {Durganandini}(2016)}]{pradeep2}%
  \BibitemOpen
  \bibfield  {author} {\bibinfo {author} {\bibfnamefont {P.}~\bibnamefont
  {Thakur}}\ and\ \bibinfo {author} {\bibfnamefont {P.}~\bibnamefont
  {Durganandini}},\ }\href {\doibase 10.1063/1.4948157} {\bibfield  {journal}
  {\bibinfo  {journal} {AIP Conference Proceedings}\ }\textbf {\bibinfo
  {volume} {1731}},\ \bibinfo {pages} {130051} (\bibinfo {year} {2016})},\
  \Eprint
  {http://arxiv.org/abs/https://aip.scitation.org/doi/pdf/10.1063/1.4948157}
  {https://aip.scitation.org/doi/pdf/10.1063/1.4948157} \BibitemShut {NoStop}%
\bibitem [{\citenamefont {Calixto}\ and\ \citenamefont
  {Romera}(2015)}]{ipr_appl}%
  \BibitemOpen
  \bibfield  {author} {\bibinfo {author} {\bibfnamefont {M.}~\bibnamefont
  {Calixto}}\ and\ \bibinfo {author} {\bibfnamefont {E.}~\bibnamefont
  {Romera}},\ }\href {\doibase 10.1088/1742-5468/2015/06/P06029} {\bibfield
  {journal} {\bibinfo  {journal} {Journal of Statistical Mechanics: Theory and
  Experiment}\ }\textbf {\bibinfo {volume} {2015}} (\bibinfo {year} {2015}),\
  10.1088/1742-5468/2015/06/P06029}\BibitemShut {NoStop}%
\bibitem [{\citenamefont {Bera}\ \emph {et~al.}(2015)\citenamefont {Bera},
  \citenamefont {Schomerus}, \citenamefont {Heidrich-Meisner},\ and\
  \citenamefont {Bardarson}}]{Bera}%
  \BibitemOpen
  \bibfield  {author} {\bibinfo {author} {\bibfnamefont {S.}~\bibnamefont
  {Bera}}, \bibinfo {author} {\bibfnamefont {H.}~\bibnamefont {Schomerus}},
  \bibinfo {author} {\bibfnamefont {F.}~\bibnamefont {Heidrich-Meisner}}, \
  and\ \bibinfo {author} {\bibfnamefont {J.~H.}\ \bibnamefont {Bardarson}},\
  }\href {\doibase 10.1103/PhysRevLett.115.046603} {\bibfield  {journal}
  {\bibinfo  {journal} {Phys. Rev. Lett.}\ }\textbf {\bibinfo {volume} {115}},\
  \bibinfo {pages} {046603} (\bibinfo {year} {2015})}\BibitemShut {NoStop}%
\bibitem [{\citenamefont {Krapivsky}\ \emph {et~al.}(2018)\citenamefont
  {Krapivsky}, \citenamefont {Luck},\ and\ \citenamefont
  {Mallick}}]{Krapivsky4}%
  \BibitemOpen
  \bibfield  {author} {\bibinfo {author} {\bibfnamefont {P.~L.}\ \bibnamefont
  {Krapivsky}}, \bibinfo {author} {\bibfnamefont {J.~M.}\ \bibnamefont {Luck}},
  \ and\ \bibinfo {author} {\bibfnamefont {K.}~\bibnamefont {Mallick}},\ }\href
  {\doibase 10.1088/1742-5468/aaa79a} {\bibfield  {journal} {\bibinfo
  {journal} {Journal of Statistical Mechanics: Theory and Experiment}\ }\textbf
  {\bibinfo {volume} {2018}},\ \bibinfo {pages} {023104} (\bibinfo {year}
  {2018})}\BibitemShut {NoStop}%
\bibitem [{\citenamefont {{Bulchandani}}\ and\ \citenamefont
  {{Karrasch}}(2018)}]{Bulchandani}%
  \BibitemOpen
  \bibfield  {author} {\bibinfo {author} {\bibfnamefont {V.~B.}\ \bibnamefont
  {{Bulchandani}}}\ and\ \bibinfo {author} {\bibfnamefont {C.}~\bibnamefont
  {{Karrasch}}},\ }\href@noop {} {\bibfield  {journal} {\bibinfo  {journal}
  {arXiv e-prints}\ ,\ \bibinfo {eid} {arXiv:1810.08227}} (\bibinfo {year}
  {2018})}\BibitemShut {NoStop}%
\end{thebibliography}%

\end{document}